\documentclass{article}

\usepackage{authblk}
\usepackage{amssymb}

\usepackage[sorting=none,citestyle=numeric-comp]{biblatex} 
\usepackage{booktabs} 

\usepackage{changepage}
\usepackage{changes}
\usepackage{chemformula} 

\usepackage{floatpag}
\usepackage{footnote}

\usepackage[letterpaper,top=2cm,bottom=2cm,left=3cm,right=3cm,marginparwidth=2cm]{geometry}
\usepackage{graphicx}

\usepackage[T1]{fontenc}

\usepackage{hyperref} 
\hypersetup{colorlinks=true,citecolor=cyan,linkcolor=blue,urlcolor=magenta}
\usepackage{cleveref} 

\usepackage{mathtools}
\usepackage[version=4,arrows=pgf-filled, textfontname=sffamily, mathfontname=mathsf]{mhchem}
\usepackage{multirow}

\usepackage{physics}
\usepackage{subcaption}
\usepackage{tablefootnote}
\usepackage{titlesec} 
\usepackage[normalem]{ulem} 
\usepackage{textcomp}

\usepackage{enumitem,amssymb}
\usepackage{setspace}

\DeclarePairedDelimiter{\ceil}{\lceil}{\rceil}

\date{\today}

\begin{document}

\title{Feasibility of accelerating homogeneous catalyst discovery with fault-tolerant quantum computers}
\author[1]{Nicole Bellonzi}
\author[1]{Alexander Kunitsa}

\author[2,3]{Joshua T. Cantin}
\author[2]{Jorge A. Campos-Gonzalez-Angulo}

\author[1]{Maxwell D. Radin}
\author[1]{Yanbing Zhou}
\author[1]{Peter D. Johnson}

\author[2,3]{Luis A. Mart\'inez-Mart\'inez}
\author[2,3]{Mohammad Reza Jangrouei}
\author[4]{Aritra Sankar Brahmachari}
\author[2,3]{Linjun Wang}
\author[2,3]{Smik Patel}

\author[1]{Monika Kodrycka}

\author[1,2,3]{Ignacio Loaiza}

\author[2,3]{Robert A. Lang}
\author[1,2]{Al\'an Aspuru-Guzik}

\author[2,3]{Artur F. Izmaylov}

\author[1]{Jhonathan Romero Fontalvo}
\author[1]{Yudong Cao}

\affil[1]{Zapata AI Inc., Boston, MA 02110 USA}
\affil[2]{Chemical Physics Theory Group, Department of Chemistry, University of Toronto, Toronto, Ontario M5S 3H6, Canada}
\affil[3]{Department of Physical and Environmental Sciences, University of Toronto Scarborough, Toronto, Ontario M1C 1A4, Canada}
\affil[4]{Indian Institute of Science Education and Research, Kolkata, West Bengal, India}

\maketitle
\thispagestyle{empty}
\begin{abstract}
The industrial manufacturing of chemicals consumes a significant amount of energy and raw materials. In principle, the development of new catalysts could greatly improve the efficiency of chemical production. However, the discovery of viable catalysts can be exceedingly challenging because it is difficult to know the efficacy of a candidate without experimentally synthesizing and characterizing it. This study explores the feasibility of using fault-tolerant quantum computers to accelerate the discovery of homogeneous catalysts for nitrogen fixation, an industrially important chemical process. It introduces a set of ground-state energy estimation problems representative of calculations needed for the discovery of homogeneous catalysts and analyzes them on three dimensions:  economic utility, classical hardness, and quantum resource requirements. For the highest utility problem considered, two steps of a catalytic cycle for the generation of cyanate anion from dinitrogen, the economic utility of running these computations is estimated to be \$200,000, and the required runtime for double-factorized phase estimation on an fault-tolerant superconducting device is estimated under conservative assumptions to be 139,000 QPU-hours. The computational cost of an equivalent DMRG calculation is estimated to be about 400,000 CPU-hours. These results suggest that, with continued development, it will be feasible for fault-tolerant quantum computers to accelerate the discovery of homogeneous catalysts.
\end{abstract}

\pagebreak
\tableofcontents

\pagebreak
\section{Introduction}
Quantum computing holds promise for advancing chemical research, particularly in catalyst discovery. This process, crucial for identifying optimal reaction pathways, is poised to benefit from the novel computational approaches to electronic structure calculations provided by quantum computing~\cite{elfving2020quantum}. 
Previous studies have assessed the efficacy of quantum algorithms for ground-state energy estimation (GSEE) across both near-term~\cite{gonthier2022,johnson2022reducing,kuehn2019accuracy} and error-corrected~\cite{vonBurg2021,babbush2018linear,Lee2020,kim2022electrolyte,reiher2017,goings2022,elfving2020quantum,berry2023quantum,rubin2023bloch,su2021first} frameworks, revealing their potential advantages over classical methods for large complex molecular systems~\cite{vonBurg2021,babbush2018linear,Lee2020,kim2022electrolyte,goings2022,elfving2020quantum}, despite the absence of evidence for an exponential asymptotic speedup~\cite{Lee2023}. However, the broader implications of quantum computing for catalyst discovery and other chemical applications remain largely unexplored. Notably, the economic benefits of quantum computational chemistry and specific performance requirements (e.g., algorithm success probability) for practical applications have yet to be clearly defined.

In the field of catalyst design, homogeneous catalysis plays a pivotal role in developing sustainable processes across multiple sectors of the chemical and pharmaceutical industries, from mass production of drugs and plastics to carbon capture and renewable energy. Although heterogeneous catalysis is traditionally favored for large-scale applications due to its robustness and ease of separation, homogeneous catalysts offer superior activity and selectivity under mild conditions. This advantage, combined with the tunable nature of transition-metal centers in these catalysts, opens up possibilities for optimizing multi-step chemical transformations, approaching the efficiency of biochemical systems. 
Despite decades of progress through classical molecular modeling methods like density functional theory (DFT) and coupled-cluster (CC) methods~\cite{ahn_design_2019}, catalyst discovery still largely depends on experimental methods with limited theoretical guidance.

This work introduces a set of GSEE instances that serve as prototypes for industrially relevant challenges with high utility. We employ a utility-driven benchmarking methodology to evaluate quantum resources for state-of-the-art Quantum Phase Estimation (QPE) algorithms, comparing them to advanced classical counterparts to assess the impact of quantum computing on homogeneous catalyst discovery and modeling kinetics and thermodynamics of small molecule activation.

A key aspect of this study is its focus on homogeneous catalysis for nitrogen fixation, a process vital for converting atmospheric nitrogen into ammonia and other useful nitrogen compounds. The quest for efficient and sustainable nitrogen fixation methods, less reliant on fossil fuels than the traditional Haber-Bosch process, underscores the need for innovative catalytic approaches that operate under milder conditions with improved efficiency and selectivity~\cite{roux_molecular_2017, tanabe_comprehensive_2021, westhead_near_2023}. This paper highlights advances in catalyst design that facilitate the activation of the challenging N$\equiv$N bond, drawing parallels with biological nitrogen fixation systems.

Our benchmarking efforts compile three realistic systems for molecular discovery in nitrogen fixation, quantifying their economic utility and the computational resources required for both classical and quantum estimation of reaction energies and barriers. The findings suggest that homogeneous catalysis is an ideal candidate for demonstrating quantum advantage, with the potential to significantly enhance our capability to design and understand catalytic processes, thereby extending the frontiers of computational chemistry.
\section{Problem Description}
\subsection{Introduction to Catalyst Design}
For a catalyst to be effective, several conditions must be met. It should produce the desired product selectively with high yield or high enantiomeric excess (for asymmetric synthesis), as well as at a high rate, measured, for example, by turnover frequency (TOF). Moreover, the catalyst should be resistant to poisoning or other degradation processes such as dimerization, auto-oxidation, and ligand dissociation. It also needs to be synthetically accessible and cost-effective for commercial viability. These challenges are compounded in multi-component catalytic systems and require a multi-scale approach that integrates atomistic simulation and microkinetic modeling. Identifying the main reaction pathways and the dominant products, whether through chemical expertise or advanced reaction network exploration tools~\cite{hashemi_renegate_2022,steiner_autonomous_2022}, is a starting point for catalyst design. This step is computationally demanding as it involves mapping out the free energy landscape by ranking multiple reaction intermediates and transition states based on their energies.

According to the transition state theory, the chemical reaction rate $k^{\ne}$ at temperature $T$ depends on the free energy difference between the reactants and the transition state, i.e., the saddle point along the minimum energy path connecting reactants to products on the ground state potential energy surface:
\begin{align}
 k^{\ne} = \frac{k_BT}{h} e^{-\frac{\Delta G_{T}^{\circ\ne}}{k_BT}},
\end{align}
where $\Delta G_{T}^{\circ\ne}$, the standard free activation energy, included the electronic ground state energy difference between the reactants and the transition state, as well as vibrational and thermal corrections. Exponential sensitivity of $k^{\ne}$ to $\Delta G_{T}^{\circ\ne}$ makes it important to calculate the latter with high accuracy. By convention, chemical accuracy is defined as an error margin within $1$ kcal/mol, which is considered acceptable for most practical applications. 
However, attaining this level of accuracy, particularly for transition metal compounds, is rarely achievable with modern DFT methods~\cite{simm_systematic_2016,vogiatzis_computational_2019}, which often struggle to meet this standard due to the complexity of accurately simulating electron interactions and other quantum mechanical effects.

Identifying elementary steps with low $k^{\ne}$ allows for pinpointing bottlenecks in the reaction network, aiding in the extraction of the core mechanism or even simplifying it to a few rate-limiting steps. Modern approaches like the energy span model~\cite{kozuch_how_2011, kozuch_refinement_2012} can be utilized here to estimate TOF and other relevant parameters for optimizing the catalyst design. 
The main ingredients of the design workflows are candidate generation and evaluation of their properties~\cite{foscato_automated_2020}.
The latter usually requires a physical model describing the stability and reactivity of the intermediates participating in the reaction.
While it is feasible to apply DFT for that purpose on a small scale, when performing knowledge-based optimization by introducing minor modifications to the existing catalyst, its computational cost is prohibitive for high-throughput screening.
Such tuning is made under the assumption of an unchanged reaction mechanism and represents the simplest mode of catalyst design.
Even then, the use of DFT is often limited to refining the energies of the key intermediates and transition states, while the task of obtaining their geometries is performed with cheaper semi-empirical~\cite{grimme_robust_2017} or force-field methods~\cite{hamza_origin_2020, hansen_prediction_2016, lin_multiconfiguration_2006}.

The downside of this approach is its inability to discover unconventional catalyst structures due to insufficient exploration of the chemical space.
More powerful high-throughput screening techniques, while addressing this issue, replace electronic structure calculations with even more severe approximations to evaluate a large number of proposed candidates.
Most of the popular strategies for mapping out larger portions of chemical space replace electronic structure calculations with statistical techniques such as scaling relations~\cite{wodrich_accessing_2016,wodrich_activity-based_2019}
or machine learning~\cite{janet_navigating_2021, nandy_computational_2021}.
Less expensive methods, however, have limited predictive power as they may not be transferable, and their quality critically depends on the reference data from which they were derived.
While high-quality training data is generally essential for robust predictive models, this requirement raises questions about the potential impact of efficient algorithms that could surpass the accuracy of DFT while incurring comparable or lower costs.

\subsection{Homogeneous Catalysis: Advantages and Specific Applications}

Homogeneous catalysis offers several advantages in terms of activity and selectivity. These catalysts often operate under milder conditions and can be effectively scaled to industrial sizes. The controlled environment allows for precise tuning of catalytic sites, leading to improved performance and minimized by-products. By modifying the ligand environment around the metal center, chemists can finely control the electronic and steric properties of the catalyst, enhancing its affinity for specific substrates and its resistance to deactivation. This precision in catalyst design is crucial for optimizing reaction pathways, reducing side reactions, and boosting overall efficiency and selectivity for industrial applications.

These advantages are particularly beneficial in nitrogen fixation, where the triple bond in dinitrogen (N$\equiv$N) presents a significant kinetic barrier to chemical conversion. Homogeneous catalysts can facilitate the activation of this bond under conditions that are less harsh than those required by the standard Haber-Bosch process. 

\subsection{Case Study in Nitrogen Fixation}
\label{sec:app_paper_prob_instances}
Advances in homogeneous catalysis for nitrogen fixation have been fueled by breakthroughs in catalyst design, including the development of complexes that can activate and reduce dinitrogen to ammonia or other nitrogenous molecules directly. These catalysts often feature transition metals that facilitate electron transfer to dinitrogen, mimicking the action of nitrogenase enzymes, which fix nitrogen under ambient conditions in biological systems.
Following the original work by Schrock et al.~\cite{yandulov_catalytic_2003} and Cummins et al.~\cite{laplaza_dinitrogen_1995}, some of the best-studied homogeneous nitrogen fixation systems are molybdenum-based. Research in this field has not only expanded the understanding of dinitrogen activation but also opened new avenues for catalytic innovation, highlighting the role of ligand design, metal coordination environments, and reaction conditions in achieving high selectivity and yields.  
Table \ref{tab:hamil_overview} provides an overview of the included Hamiltonians for each reaction. Molecular labels are based on the original investigations with minimal adjustments. 

\begin{table}[!p]
    \setlength\tabcolsep{0pt}
    \centering
    \caption{Nitrogen Fixation Hamiltonians. Each unique molecule and active space can be identified with the Molecule ID and the number of orbitals $N_o$.  Here, we also include the number of electrons $N_e$, the charge, and the multiplicities of the Hamiltonians. Finally, we report the size of the Hilbert space, representing the complexity of the quantum state space.}\label{tab:hamil_overview}
    \begin{subtable}{0.5\textwidth}
    \caption{Schrock Catalyst  $\text{Mo} \text{N}_2$ (Section \ref{sec:schrock_mo_n2_instance})} \label{tab:hamils_mo_n2}
    \begin{tabular*}{\textwidth}{@{\extracolsep{\fill}}l*{5}{c}}
        \toprule
        \shortstack{Molecule\\ID} & {$N_o$} & {$N_e$} & {Charge} & \shortstack{Multi-\\plicity} & \shortstack{Hilbert\\Space (log)} \\ \midrule
        MoN$_2$ & 33 & 45 & 0 & 2 & 16\\ 
        MoN$_2^-$ & 33 & 46 & -1 & 1 & 15\\ 
        Fe(Cp)$_2$ & 46 & 58 & 0 & 1 & 24\\ 
        Fe(Cp)$_2^+$ & 46 & 57 & 1 & 2 & 24\\ 
        \bottomrule
    \end{tabular*}
    \end{subtable}
    
    \vspace*{5mm}
    
    \begin{subtable}{0.5\textwidth}
    \caption{Bridged Dimolybdenum  $\text{Mo}_2 \text{N}_2$ (Section \ref{sec:mo2_n2_instance})}\label{tab:hamils_1}
    \begin{tabular*}{\textwidth}{@{\extracolsep{\fill}}l*{5}{c}}
        \toprule
        \shortstack{Molecule\\ID} & {$N_o$} & {$N_e$} & {Charge} & \shortstack{Multi-\\plicity} & \shortstack{Hilbert\\Space (log)} \\ \midrule
        1-Lut$_{Re}$ & 69 & 90 & 1 & 1 & 36\\
        1-Lut$_{TS}$ & 70 & 90 & 1 & 1 & 37\\ 
        II-Lut$_{Prod}$ & 70 & 90 & 1 & 1 & 37\\ 
        \bottomrule
    \end{tabular*}
    \end{subtable}
    
    \vspace*{5mm}
    
    \begin{subtable}{.5\textwidth}
    \caption{$\text{Mo} {\text{N}_2}$ Pincer reactions  (Section \ref{sec:mo_n2_pincer_instance})} \label{tab:hamilss_mo_n2_pincer}
    \begin{tabular*}{\textwidth}{@{\extracolsep{\fill}}l*{5}{c}}
        \toprule
        \shortstack{Molecule\\ID} & {$N_o$} & {$N_e$} & {Charge} & \shortstack{Multi-\\plicity} & \shortstack{Hilbert\\Space (log)} \\ \midrule
        \multicolumn{6}{c}{(step i) smaller active space} \\
        \midrule
        RC & 32 & 50 & 0 & 1 & 13\\
        TS$_{1/2}$ & 27 & 44 & 0 & 1 & 9\\
        PC & 32 & 50 & 0 & 1 & 13\\
        2 & 33 & 50 & 0 & 1 & 14\\ 
        \midrule
        \multicolumn{6}{c}{(step i) larger active space}\\
        \midrule
        RC & 51 & 74 & 0 & 1 & 24\\
        TS$_{1/2}$ & 51 & 74 & 0 & 1 & 24\\ 
        PC & 51 & 74 & 0 & 1 & 24\\ 
        2 & 52 & 74 & 0 & 1 & 25\\ 
        \midrule
        \multicolumn{6}{c}{(step ii) smaller active space}\\
        \midrule
        I & 56 & 77 & -1 & 2 & 28\\
        TS$_{\text{I/4a}}$ & 56 & 75 & -1 & 4 & 28\\
        PC$^-$ & 55 & 75 & -1 & 4 & 27\\
        4a & 24 & 39 & 0 & 4 & 8\\
        \midrule
        \multicolumn{6}{c}{(step ii) larger active space}\\
        \midrule
        I & 75 & 101 & -1 & 2 & 39\\
        TS$_{\text{I/4a}}$ & 75 & 99 & -1 & 4 & 39\\
        PC$^-$ & 73 & 99 & -1 & 4 & 37\\
        4a & 43 & 63 & 0 & 4 & 19\\
        \bottomrule
    \end{tabular*}
    \end{subtable}
\end{table}

\subsubsection{The Schrock Catalyst}
\label{sec:schrock_mo_n2_instance}

\begin{figure}[ht]
    \centering
    \includegraphics[width=\textwidth]{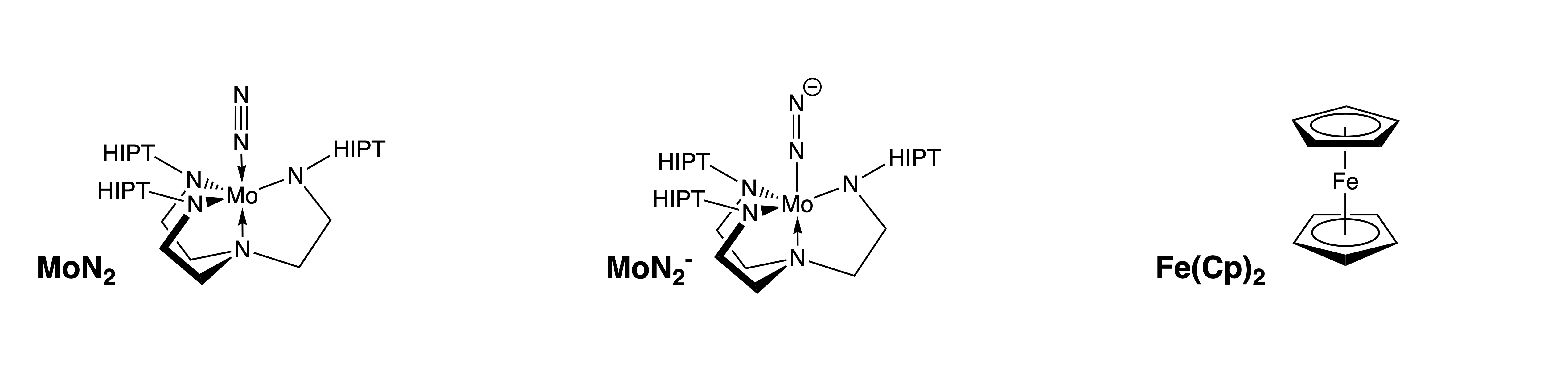}
    \caption{Molecules included for the Schrock catalyst, focused on the first step of the catalytic cycle, in which the base catalyst MoN$_2$ is reduced to MoN$_2^{-}$ by ferrocene Fe(Cp)$_2$.}\label{fig:chem_mo_n2}
\end{figure}

One of the first selective molybdenum (Mo) complexes to reduce dinitrogen to ammonia with minimal side products was developed by Schrock and colleagues~\cite{yandulov_catalytic_2003}. This catalyst demonstrated the remarkable ability to reduce dinitrogen to ammonia under ambient conditions, requiring only an electron and proton source. Optimal catalytic activity was achieved with a hexaisopropyl terphenyl substituent (HIPT), striking a balance between steric hindrance at the metal center to prevent dimerization and accessibility to the active site.

Our instance is based on the work by Schenk and colleagues~\cite{shrock_2008}, whose computational studies revealed that simplified model systems often fall short of capturing the intricacies of the catalytic cycle, highlighting the need for advanced quantum chemical approaches capable of describing realistic catalysts. The incorporation of sizable chelate ligands, HIPT, not only influences steric effects but also significantly alters the electronic structure at the reaction center, thereby impacting the overall thermodynamics of the process. This instance focuses on the initial nitrogen reduction step by the Schrock catalyst in the presence of ferrocene, $\text{Fe(Cp)}_2$, as a typical reducing agent, shown in Figure \ref{fig:chem_mo_n2}.

\subsubsection{Bridged Dimolybdenum Complex}
\label{sec:mo2_n2_instance}

\begin{figure}[ht]
    \centering
    \includegraphics[width=\textwidth]{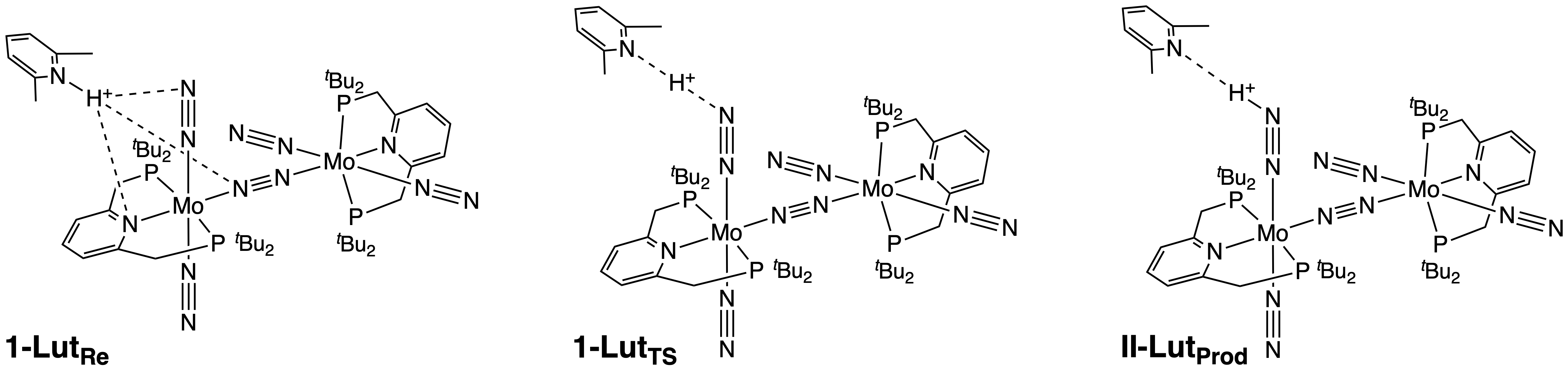}
    \caption{Molecules included for the Bridged Dimolybdenum Complex, focused on the initial protonation step in which a terminal nitrogen ligand of $1$-Lut$_{Re}$ is protonated to II-Lut$_{Prod}$ through transition state $1$-Lut$_{TS}$.}\label{fig:chem_lut}
\end{figure}

After the Schrock catalyst, much work focused on mononuclear molybdenum complexes for nitrogen fixation. In efforts to improve these catalysts, scientists developed a dinuclear molybdenum complex that improved efficiency from less than 8 equivalents of ammonia synthesized per molybdenum atom to greater than 11. Tanaka and coworkers~\cite{Tanaka2014} explored the energy landscape of various reaction pathways and determined that a terminal dinitrogen ligand undergoes the first protonation step in the catalytic cycle, not the more strongly activated bridging dinitrogen ligand, due to the steric protection of the nitrogen bridge by the chelate ligands.
Our instance explores the initial protonation step on the terminal dinitrogen ligand, including the reactant state (1-Lut$_{Re}$), transition state (1-Lut$_{TS}$), and terminally protonated product state (II-Lut$_{Prod}$), shown in Figure \ref{fig:chem_lut}.

\subsubsection{Molybdenum Pincer}
\label{sec:mo_n2_pincer_instance}

\begin{figure}[ht]
    \centering
    \includegraphics[width=\textwidth]{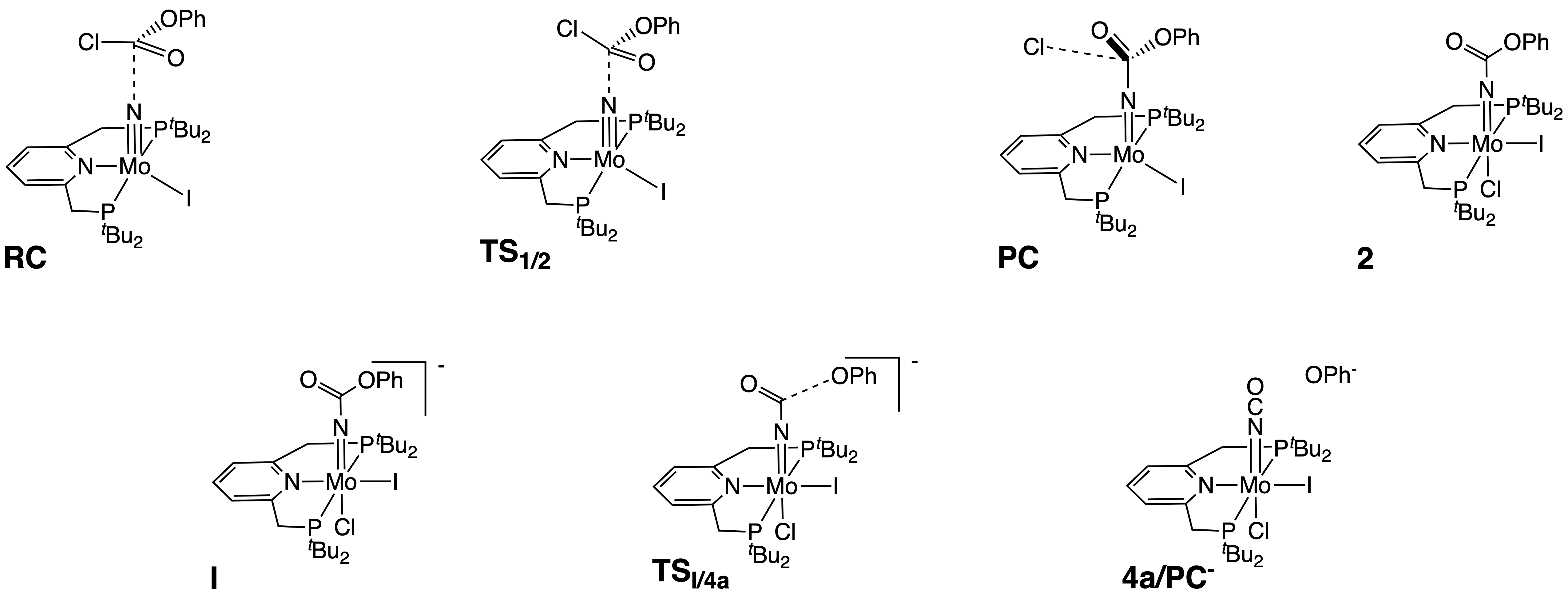}
    \caption{Molecules included for the Molybdenum Pincer Complex, focused on the two steps of the catalytic cycle. During step (i), the reactant complex RC transforms into PC via TS$_{1/2}$, following which the released Cl$^-$ migrates to its ultimate position, forming $2$. 
    In step (ii), the reduction of complex $2$ initiates the cleavage of the C--OPh bond in complex I, leading to the formation of PC$^-$ via TS$_{I/4a}$. PC$^-$ includes complex 4a and a released OPh$^-$.}\label{fig:chem_pincer}
\end{figure} 

While much of nitrogen fixation research focuses on generating ammonia, there are other useful products, such as cyanate anions [N$\equiv$ C--O]$^-$. Itabashi and coworkers~\cite{Itabashi2022} studied the process of employing molybdenum complexes featuring a pyridine-based PNP-type pincer ligand under standard reaction conditions. They identified and characterized key intermediates, including the reactant complex (RC), transition state (TS), and product complex (PC), for various steps of the cycle, shedding light on the mechanistic intricacies underlying this synthetic pathway.

In our instance, we showcase two steps of the catalytic cycle for generating cyanate anion NCO from dinitrogen. These steps, denoted as (i) and (ii) in their work, are characterized by high spin states. Step (i) starts with an approach of phenyl chloroformate to the base catalyst, together forming the RC. Through TS$_{1/2}$ the carbonyl carbon atom forms a bond with the nitride ligand, cleaving the C--Cl bond, leading to the PC. The Cl$^-$ anion, upon its release, takes up the unoccupied coordination site on the $Mo$ center, resulting in $2$. 
Step (ii) begins with the reduced state of complex $2$, referred to as complex I. The C--OPh bond within I (doublet) undergoes cleavage via a spin inversion at TS$_{I/4a}$ (quartet), resulting in the formation of PC$^-$ (quartet), comprising complex 4a and a free phenolate anion (OPh$^-$). To explore a variety of problem sizes, we include both small and large active spaces for this problem instance.

\subsection{Active Space Selection}
We applied the atomic valence active space (AVAS)~\cite{sayfutyarova_automated_2017} method to the corresponding Hartree-Fock solutions to define the Hamiltonians for chemical moieties participating in the nitrogen fixation reactions. AVAS automatically identifies orbitals with significant overlap with predefined atomic valence orbitals, ensuring that the resulting active space is relevant to the chemical process or properties of interest. 

Within the AVAS formalism, the target atomic orbitals are expressed in an auxiliary minimum basis. The projection operator on the space they span is diagonalized separately within occupied and virtual molecular orbital subspaces. Subsequently, the eigenvectors with sufficiently large eigenvalues are chosen as the basis of the active space. The eigenvalue threshold is the only numerical parameter set by a user. Once the active orbitals are selected, the rest are marked as doubly occupied or virtual, splitting the original molecular orbitals into three subsets. If necessary, the orbitals can be canonicalized separately within each subset.

Throughout this work, we used STO-$3$G as a minimum basis for AVAS and applied the default projection threshold of $0.2$ as implemented in PySCF~\cite{Sun2020}. For all systems, only the alpha orbital subset was included in the active space selection, which is the default setting in PySCF.
The orbitals were not canonicalized. Detailed specifications of the atomic valence spaces for each system are included in the Supplementary Information.  

\subsection{The Path to Better Nitrogen Fixation Catalysts}
These problem instances exemplify the main model chemistries studied in the context of early homogeneous nitrogen fixation. Later catalyst redesign efforts were aimed at improving their performance in several aspects, including (a) Faradic efficiency (the portion of electrons from the reducing agent going toward making the desired product), (b) turnover number (TON; a measure of catalyst stability), and (c) TOF (a measure of reaction rate). However, despite these efforts, none of the currently known synthetic systems outperforms nitrogenase across all metrics. This underscores the need for advancements in computational methods to achieve higher accuracy in predicting molecular energies, which could significantly enhance the design and optimization of homogeneous catalysts.
Some of the best catalysts in the class of homogeneous transition metal systems are molybdenum complexes with PCP-type~\cite{ashida_catalytic_2023} or PNP-type~\cite{ashida_molybdenum-catalysed_2019} pincer ligands, discovered with the aid of DFT calculations, and in some cases through direct tuning of the ligand properties~\cite{ashida_catalytic_2023}. However, despite impressive selectivity and activity, the latest generation of nitrogen fixation catalysts~\cite{chalkley_catalytic_2020} falls short of reaching the stability required for industrial applications~\cite{westhead_near_2023}.

The desire to decarbonize ammonia production and the decreasing price of solar and wind energy~\cite{lcoe_2023_lazard} stimulated research into alternative nitrogen fixation methods that utilize sustainable energy sources to perform nitrogen cleavage. This can be accomplished in electrocatalytic systems, where protons and electrons can be supplied, for example, by water oxidation reaction~\cite{seh_combining_2017} driven by solar energy. 
To make such processes economically viable, new electrode materials must be designed to enable nitrogen fixation at low overpotential and high current density. 
Additionally, side reactions such as hydrogen evolution need to be suppressed or controlled~\cite{singh_electrochemical_2017}. This is a challenge much harder to address in heterogeneous systems compared to their homogeneous counterparts where ligands can be engineered to limit hydrogen access to the reaction center. Fundamentally new design principles for such systems have been put forward by several research 
groups~\cite{singh_electrochemical_2017,pedersen_dual-metal_2022, yang_surface_2023}. However, theoretically predicted optimal electrode materials are often hard to synthesize and test under realistic operating conditions. The only class of electrochemical systems that emerged as a lead contender for ambient nitrogen fixation is the one where ammonia production is lithium-mediated~\cite{westhead_near_2023}. The detailed mechanism of the process and the role of lithium, however, remain the area of active research and likely hold the key to further technological advancements~\cite{westhead_is_2021}.    

\section{Classical Methods and Required Resources}
This section discuses state-of-the-art classical ground state energy estimation methods and the limits to their performance. 
By identifying these performance limits, we can establish thresholds that new methods, whether quantum or classical, must exceed to provide additional utility. Methods discussed below include 
Coupled Cluster (CC) and the Density Matrix Renormalization Group (DMRG).
We focused on methods that can be systematically improved towards a full configuration computation, as that is most similar to what a quantum computer will compute.

Density Functional Theory (DFT) is widely employed for predicting material properties, primarily due to its lower computational cost compared to wave function theories like CC and DMRG. While DFT is viable for analyzing large catalysts on classical computers, its accuracy is generally lower and can vary significantly with the specific case and properties under study~\cite{VERMA2020302}. The reliability of DFT is limited by the use of approximate model functionals~\cite{Cohen2012}, as the exact exchange-correlation functional remains unknown. This is particularly problematic in the strong correlation regime, where DFT's single-reference nature struggles to account for static correlations described by a small number of nearly degenerate configurations. Although the multiconfiguration pair-density functional theory (MC-PDFT) framework has been developed to address these shortcomings~\cite{Li2014}, it only partially mitigates the fundamental limitations of DFT. Moreover, DFT is plagued by self-interaction error~\cite{Zhang1998}, where an electron unphysically interacts with itself, and additional inaccuracies such as the delocalization error~\cite{Yang2008}, which is well defined in terms of fractional charge and spin~\cite{YangWeitao2009}. Despite several advancements in modern density functionals~\cite{Narbe2017} to address these issues, their effectiveness is inconsistent, and predicting the success or failure of a specific functional in complex electronic structures remains uncertain. This intrinsic limitation contrasts sharply with the systematic improvements achievable in methods like CC and DMRG.

\subsection{Coupled Cluster}
The numerical technique known as Coupled Cluster~\cite{coester1960, cizek1966, paldus1977, bartlett1978} aims to construct multi-electron wavefunctions from the Hartree-Fock (HF) molecular orbitals, accounting for electron correlation with an exponential cluster operator. This approach ensures size extensivity~\cite{Bartlett2007}. The CC methods have been successful at predicting accurate energetics of small molecules, and they can be used to estimate properties of materials~\cite{zhang2019}. Prominent examples of its successes can be found in the calculation of excitation energies, vibronic spectra, Raman spectral intensities, and valence-bond solutions, among others~\cite{Sekino1984, Perera1999, Cullen1996, Bishop1991}. Indeed, its version considering single and double excitations with perturbative corrections for triple excitations [CCSD(T)] is broadly recognized as the ``gold standard'' of computational chemistry. Although the method scales polynomially with the system size, as opposed to exponentially, it still remains challenging to apply it to systems beyond 15-20 atoms~\cite{Datta2021}, and its use is typically restricted to the generation of benchmark sets to calibrate other less expensive methods such as DFT. In this work, we explore the capabilities of CCSD and CCSD(T) to yield accurate solutions for a set of reference molecules and attempt to extend diagnostic tools to instances of industrially relevant homogeneous catalysts.

\subsubsection{Software and Implementation}

We extract the relevant quantities from calculations at the HF, CCSD, CCSD(T), and FCI levels of theory run on the Python-based Simulations of Chemistry Framework (PySCF)~\cite{Sun2018,Sun2020}. Furthermore, we analyze available FCI solutions with the ClusterDec suite~\cite{Lehtola2017}.

\subsubsection{Method Performance}

Table~\ref{tab:cc_all_results} shows results for CCSD and CCSD(T) calculations of each of the Hamiltonians from the catalytic cycles of the Schrock, Bridged Dimolybdenum and Molybdenum Pincer problem instances (see Section~\ref{sec:app_paper_prob_instances}). 

\begin{table}[hp]
    \setlength\tabcolsep{0pt}
    \centering
    \caption{CCSD and CCSD(T) Results for the nitrogen fixation catalysts. The red-highlighted quantities indicate non-converged energy.}\label{tab:cc_all_results}
    \begin{subtable}{\textwidth}
    \caption{CC results for Schrock catalyst (Section~\ref{sec:schrock_mo_n2_instance})} \label{tab:cc_results_mo_n2}
    \begin{tabular*}{\textwidth}{@{\extracolsep{\fill}}l*{5}{c}}
        \toprule
        \shortstack{Molecule ID}  & $N_o$ &  \shortstack{CCSD \\energy \\ ($E_h$)} & \shortstack{CCSD\\ CPU time \\ (s)} & \shortstack{CCSD(T)\\correction \\ ($\times 10^{-3}E_h$)} & \shortstack{CCSD(T) \\ CPU time \\ (s)} \\ \midrule
        MoN$_2^-$ & 33 & -8694.2836 & 0.44 & -5.0222 & 0.06\\  
        MoN$_2$ & 33 & -8694.0684 & 3.46 & -4.9397 & 0.3 \\  
        Fe(Cp)$_2^+$ & 46 & {\color{red}-1647.0148} & 576.76 & -14.3747 & 3.1 \\  
        Fe(Cp)$_2$ & 46 & -1647.3132 & 2.66 & -12.7239 & 0.53 \\  
        \bottomrule
    \end{tabular*}
    \end{subtable}
    
    \vspace*{5mm}

    \begin{subtable}{\textwidth}
    \caption{CC results for Bridged Dimolybdenum (Section~\ref{sec:mo2_n2_instance})}\label{tab:cc_results_1}
    \begin{tabular*}{\textwidth}{@{\extracolsep{\fill}}l*{5}{c}} 
        \toprule
        \shortstack{Molecule ID}  & $N_o$ &  \shortstack{CCSD \\energy \\ ($E_h$)} & \shortstack{CCSD\\ CPU time \\ (s)} & \shortstack{CCSD(T)\\correction \\ ($\times 10^{-3}E_h$)} & \shortstack{CCSD(T) \\ CPU time \\ (s)} \\ \midrule
        1-Lut$_{Re}$ &  69 & -12049.7828 & 17.63 & -15.4600 & 7.63 \\
        1-Lut$_{TS}$ &  70 & -12049.7956 & 20.67 & -15.1356 & 8.72 \\
        II-Lut$_{Prod}$ &  70 & -12049.8323 & 20.74 & -13.9025 & 8.84 \\
        \bottomrule
    \end{tabular*}
    \end{subtable}
    
    \vspace*{5mm}
    
    \begin{subtable}{\textwidth}
    \caption{CC results for Mo Pincer reactions (Section~\ref{sec:mo_n2_pincer_instance})} \label{tab:cc_results_mo_n2_pincer}
    \begin{tabular*}{\textwidth}{@{\extracolsep{\fill}}l*{5}{c}}
        \toprule
        \shortstack{Molecule ID}  & $N_o$ &  \shortstack{CCSD \\energy \\ ($E_h$)} & \shortstack{CCSD\\ CPU time \\ (s)} & \shortstack{CCSD(T)\\correction \\ ($\times 10^{-3}E_h$)} & \shortstack{CCSD(T) \\ CPU time \\ (s)} \\ \midrule
        \multicolumn{6}{c}{step (i) smaller active space} \\
        \midrule
        RC & 32 & -5412.0320 & 0.3 & -4.5276 & 0.03 \\  
        TS$_{1/2}$ & 27 & -5413.3505 & 0.16 & -4.3792 & 0.01 \\ 
        PC & 32 & -5413.1827 &  0.27 & -4.3710 & 0.03 \\ 
        2 & 33 & -5411.9917 &  0.42 & -6.0725 & 0.04 \\  
        \midrule
        \multicolumn{6}{c}{step (i) larger active space}\\
        \midrule
        RC & 51 & -5412.0818 & 2.05 & -5.0899 & 0.8 \\  
        TS$_{1/2}$ & 51 & -5413.4594 & 2.47 & -6.1738 & 0.78 \\
        PC & 51 & -5413.2286 & 3.03 & -5.0954 & 0.8  \\ 
        2 & 52 & -5412.0368 &  3.49 & -6.2267 & 0.98 \\ 
        \midrule
        \multicolumn{6}{c}{step (ii) smaller active space}\\
        \midrule
        I &  56 & {\color{red}-5411.4326} & 1423.83 & -9.5501 & 8.72 \\ 
        TS$_{\text{I/4a}}$ &  56 & -5409.6278 & 904.61 & -14.1437 & 10.25 \\ 
        PC$^-$ &  55 & {\color{red}-5409.5052} & 1291.83 & -12.3709 & 8.29 \\
        4a & 24 & -5103.5538 & 2.1 & -2.4389 & 0.02 \\ 
        \midrule
        \multicolumn{6}{c}{step (ii) larger active space}\\
        \midrule
        I &  75 & -5411.4889 & 684.72 & -10.9233 & 73.54 \\
        TS$_{\text{I/4a}}$ &  73 & {\color{red}-5409.9698} & 1094.91 & -11.8164 & 59.62 \\ 
        PC$^-$ &  73 & -5409.6468 & 4593.95 & -12.3676 & 58.29 \\ 
        4a & 43 & -5103.8843 & 11.38 & -2.5960 & 1.22 \\
        \bottomrule
    \end{tabular*}
    \end{subtable}
\end{table}
\subsection{DMRG}\label{sec:dmrg}
The density matrix renormalization group \cite{whiteDmrg1992,whiteDmrg1993,schollwoeckDmrg2005,olivaresBlock2015,szalayDmrg2015,baiardiDmrg2020,catarinaDmrg2023,yanaiDmrg2015} (DMRG) is a variational method used in both chemistry and physics to determine various properties of many-body quantum systems, including the ground state of molecules. DMRG has been applied to a large variety of molecules, from arenes \cite{olivaresBlock2015} to dimers with large static correlation \cite{olivaresBlock2015,moritzChromium2005} to organometallic compounds \cite{olivaresBlock2015,phungPorphyrins2023}, and is now a reference method for large systems with large static correlation \cite{baiardiDmrg2020}. Here, we apply DMRG to the nitrogen fixation catalysts described in Section \ref{sec:app_paper_prob_instances}.

\subsubsection{Software and Implementation}
We use the recently developed and highly optimized \verb|Block2| open source software package \cite{zhai2023block2} for these calculations. We take advantage of various features included in \verb|Block2|, such as SU(2) symmetry and perturbative noise.

As part of determining the ground state energy from DMRG, we employ an extrapolation technique. Many different extrapolation methods have been used in the literature \cite{szalayDmrg2015}. We use a series of calculations of different bond dimension and then estimate the ground state energy by extrapolating the DMRG energy $E_{\mathrm{DMRG}}$ as a function of the truncated weight $\delta \epsilon$ via a linear fit~\cite{schollwoeckDmrg2005}:
\begin{align}
    E_{\mathrm{DMRG}} = E_{\mathrm{est}} + m\cdot\delta \epsilon \label{eqn_dmrg_extrapolation}
\end{align}
where $E_{\mathrm{est}}$ is the intercept and $m$ the slope of the linear fit. We then obtain a 95\% confidence interval for the estimated energy $E_{\mathrm{est}}$ by using the standard error of the intercept, obtainable from the fit.

After obtaining $E_{\mathrm{est}}$, we also extrapolate the bond dimension to obtain an estimate of the bond dimension $D_{\mathrm{est}}$ required for the DMRG energy to be within 1 milliHartree of $E_{\mathrm{est}}$. $D_{\mathrm{est}}$ is typically expected to be larger than the largest bond dimension used so far in the calculations, but can be less for  systems where the calculations are already well-converged to  $E_{\mathrm{est}}$. 

To perform this bond dimension extrapolation, we first perform a linear fit \cite{chan2002highly}:
\begin{align}
    \ln (E_{\mathrm{DMRG}} - E_{\mathrm{est}}) = a + b(\ln D)^2
\end{align}
Once the $a$ and $b$ parameters in the above equation are determined, the estimated bond dimension to obtain the DMRG energy within the accuracy $\delta$ of $E_{\mathrm{est}}$ is given by $D_{\mathrm{est}} = \exp{\sqrt{\frac{\ln{\delta} - a}{b}}}$. Assuming the standard errors of $a$ and $b$ are $\sigma_a$ and $\sigma_b$, respectively, we can define the four values $D_{\mathrm{est}}^{\pm\pm} = \exp{\sqrt{\frac{\ln{\delta} - (a\pm1.96\cdot\sigma_a)}{(b\pm1.96\cdot\sigma_b)}}}$. These values illustrate the variability in $D_{\mathrm{est}}$ due to fitting errors and we can determine an interval for $D_{\mathrm{est}}$ by taking the minimum and maximum of these four values.

For each Hamiltonian, we perform a series of calculations with increasing bond dimension and monitor the obtained energy $E_{\mathrm{DMRG}}$ after each new calculation, stopping the series when the change in estimated energy is sufficiently small, or a predefined limit on the bond dimension, calculation time, or RAM used is reached. For calculations run on Amazon Web Services cloud computing instances, we perform a reverse schedule in addition to the forward approach mentioned above.

The compute instances used on Amazon Web Services are \verb|c5.9xlarge| with 36 Intel Intel Xeon Platinum 8000 series ``Skylake'' 3.6 GHz cores and 72 GiB of RAM and are managed via the Orquestra\textsuperscript{\textregistered} platform provided by Zapata AI. The second computing system we use is the Niagara cluster hosted by SciNet \cite{ponce2019deploying,loken2010scinet}, who are  partnered with Compute Ontario and the Digital Research Alliance of Canada. Each node of Niagara  has 40 Intel ``Skylake'' 2.4 GHz cores or 40 Intel ``CascadeLake'' 2.5 GHz cores and has 188 GiB of RAM.  Which problem instances were run on which computing system is specified in the tables.

\subsubsection{Algorithm Performance}

\begin{figure}[tp]
    \centering
    \includegraphics[width=0.8\linewidth]{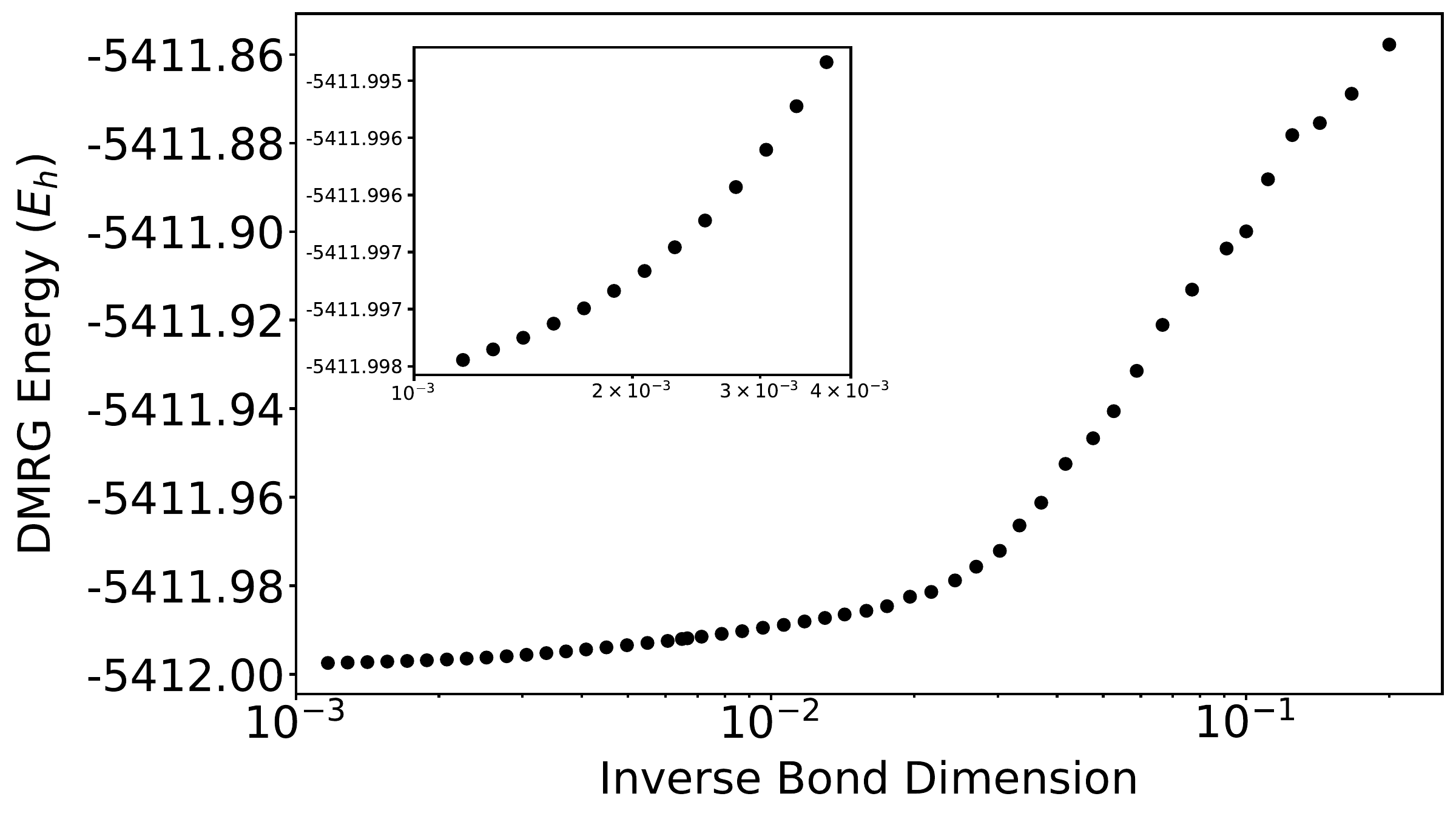}
    \caption{Example DMRG convergence plot, for Mo Pincer complex 2 using the smaller active space. Inset is a zoomed-in version of the main axes. Total wall time for these calculations on the Niagara computer cluster is about 4 hrs.}
    \label{fig:dmrg_convergence_example}
\end{figure}

Table \ref{tab:dmrg_all_results} shows the results for DMRG calculations of the nitrogen fixation catalysts mentioned in Section \ref{sec:app_paper_prob_instances}. Several molecules have both a large and a small active space version. This table contains both the DMRG energies obtained and bond dimensions reached, as well as extrapolations of the energies and bond dimensions.
An example convergence diagram for the forward approach is shown in Figure \ref{fig:dmrg_convergence_example} for Mo Pincer complex 2 using the smaller active space. 

There are some trends in the data that we can identify. Generally, the required CPU time increases with the number of orbitals in the active space (Pearson correlation coefficient, PCC, of 0.81); this is in line with the expectation that the time of a DMRG calculation varies as $N_o^4$ \cite{Wouters2014}. However, the size of the extrapolated bond dimension required to be within 1 mHa of the extrapolated energy does not correlate well with the number of orbitals (both the Pearson and Spearman correlation coefficients are low). This seems reflective of the required bond dimension being more directly a function of the entanglement structure of the ground state, rather than of the system size \cite{catarinaDmrg2023}.

Interestingly, while the transition state TS$_{I/4a}$ requires a larger bond dimension than its corresponding reactant, product, or intermediate species, the opposite is true for TS$_{1/2}$ and 1-Lut$_{TS}$. While the reason for this is not clear, one path forward would be to investigate the entanglement structure of the systems and the locality of the molecular orbitals.

As with any attempt to determine the present capabilities of a method, there will always be ways to improve the estimates made. The largest source of potential improvement would be to include orbital ordering for the DMRG calculations \cite{olivaresBlock2015}. Another source would be to add ``post-DMRG'' methods that help to account for dynamic correlation \cite{larsson2022}. After this, larger bond dimensions could be reachable by using a distributed-memory implementation of DMRG \cite{zhai2021}. Different choices for the active spaces of each of the molecules could also be made; sub-optimal active space selection is a possible  reason for the likely unrealistically large extrapolated bond dimensions for Fe(Cp)$_2^+$ and Fe(Cp)$_2$. The calculations could also be optimized with respect to sweep schedules. 

\begin{savenotes}
\begin{table}[hp]
    \setlength\tabcolsep{0pt}
    \centering
    \caption{
    DMRG Results for the nitrogen fixation catalysts. Error of the energy extrapolation~\cite{dmrg_energy_extrapolation_note}
    is a 95\% CI, with the following notation: $-5413.4663(11)=-5413.4663\pm0.0011$. ${}^\dagger$CPU Calculation Time was normalized to 20 DMRG sweeps.} \label{tab:dmrg_all_results}
    \begin{subtable}{\textwidth}
    \centering
    \caption{DMRG results for Schrock catalyst (Section \ref{sec:schrock_mo_n2_instance}), Niagara computer cluster} \label{tab:dmrg_results_mo_n2}
    \begin{tabular*}{\textwidth}{@{\extracolsep{\fill}}lcccccc}
        \toprule
        \shortstack{Molecule ID}  & $N_o$ & \shortstack{Bond\\Dimension} &  \shortstack{Energy\\($E_h$)} & \shortstack{CPU \\Calculation \\Time (hr)${}^\dagger$} &\shortstack{Extrapolated\\Energy ($E_h$)}&\shortstack{Extrapolated\\Bond Dimension\\ $[\mathrm{min},\mathrm{max}]$ } \\ \midrule 
        MoN$_2^-$ &33&1558&$-8694.2886$&91.68& $-8694.2897(1)$ &  1655 [1519, 1805]\\ 
        MoN$_2$&33&878&$-8694.0684$&30.72& $-8694.0759(7)$ &  6502 [6090, 6949]  \\  
        Fe(Cp)$_2^+$&46&599&$-1647.0880$&39.12& $-1647.1110(23)$ & 8413 [7301, 9735] \\ 
        Fe(Cp)$_2$~\cite{ferrocene_reverse_note} 
        &46&544&$-1647.3037$& 27.84& $-1647.3222(22)$ & {\color{orange} 13133 [11589, 14930]} \\  
        \bottomrule
    \end{tabular*}
    \end{subtable}
    
    \vspace*{5mm}

    \begin{subtable}{\textwidth}
    \caption{DMRG results for Bridged Dimolybdenum (Section \ref{sec:mo2_n2_instance}), AWS cloud computing}\label{tab:dmrg_results_mo2_n2}
    \begin{tabular*}{\textwidth}{@{\extracolsep{\fill}}lcccccc}
        \toprule
        \shortstack{Molecule ID}  & $N_o$ & \shortstack{Bond\\Dimension} &  \shortstack{Energy\\($E_h$)} & \shortstack{CPU \\Calculation \\Time (hr)${}^\dagger$} &\shortstack{Extrapolated\\Energy ($E_h$)}&\shortstack{Extrapolated\\Bond Dimension\\ $[\mathrm{min},\mathrm{max}]$ } \\ \midrule
        1-Lut$_{Re}$ &  69 &             450 & $-12049.6523$ &           203.35& $-12049.6675(15)$ & 2565 [1684, 3446] \\ 
        1-Lut$_{TS}$ &  70 &             420 & $-12049.6495$ &           179.79& $-12049.6574(8)$ & 1098 [1017, 1179] \\  
        II-Lut$_{Prod}$ &  70 &              420 & $-12049.6887$ &           178.74& $-12049.6973(15)$ &  1254 [894, 1614] \\  
        \bottomrule
    \end{tabular*}
    \end{subtable}
    
    \vspace*{5mm}

    \begin{subtable}{\textwidth}
    \caption{DMRG results for Mo Pincer reaction (Section \ref{sec:mo_n2_pincer_instance}), Various compute}
    \begin{tabular*}{\textwidth}{@{\extracolsep{\fill}}lcccccc}
    \toprule
        \shortstack{Molecule ID}  & $N_o$ & \shortstack{Bond\\Dimension} &  \shortstack{Energy\\($E_h$)} & \shortstack{CPU\\Calculation \\Time (hr)${}^\dagger$} &\shortstack{Extrapolated\\Energy ($E_h$)}&\shortstack{Extrapolated\\Bond Dimension\\ $[\mathrm{min},\mathrm{max}]$ } \\ \midrule %
        
        \multicolumn{7}{c}{step (i) smaller active space, Niagara computer cluster} \\ \midrule
        RC & 32  & 437 &  $-5412.0367$ & 3.30   &{$-5412.0373(1)$}&{299 [202, 457]}  \\  
        TS$_{1/2}$ & 27  & 154 & $-5413.3549$  & 0.41  &$-5413.3552(1)$&{92 [82, 103]} \\
        PC & 32  & 449 &  $-5413.1870$ &  4.51  & $-5413.1876(1)$ & 332 [315, 349] \\ 
        2 & 33  & 856 & $-5411.9974$  &  23.93  &$-5411.9980(1)$&583 [550, 619]  \\   
        
        \midrule \multicolumn{7}{c}{step (i) larger active space, Niagara computer cluster} \\ \midrule
        RC & 51  & 583 & $-5412.0839$  & 43.68  &$-5412.0898(8)$& 4126 [3777, 4516] \\ 
        TS$_{1/2}$ & 51  & 583 &  $-5413.4598$ & 51.60  &$-5413.4663(11)$&4076 [3819, 4354]    \\ 
        PC & 51  & 481 & $-5413.2293$  &  26.40 &$-5413.2366(6)$& 4888 [4638, 5156]  \\  
        2 & 52  & 530 & $-5412.0364$  &  42.96  &$-5412.0446(3)$&{ 6434 [6089, 6803]}  \\  
        
        \midrule \multicolumn{7}{c}{step (ii) smaller active space, AWS cloud computing} \\ \midrule
        I &  56 &             450 &  $-5411.4060$ &           99.39&  $-5411.4111(11)$ &   939 [743, 1135] \\  
        TS$_{\text{I/4a}}$ &  56 &            450 &  $-5409.5595$ &           109.21&  $-5409.5751(33)$ & 2454 [2052, 2856] \\  
        PC$^-$ &  55 &              450 &  $-5409.4560$ &           97.40&  $-5409.4660(33)$ &  2279 [544, 4014] \\  
        4a &  24 &              325 &  $-5103.5570$ &             0.82&  {$-5103.5570(*)$}\cite{4a_small_uncertainty_note}
        &        15 [3, 27] \\ 
        
        \midrule \multicolumn{7}{c}{step (ii) larger active space, AWS cloud computing} \\ \midrule
        I &  75 &            400 &  $-5411.4342$ &           217.05&  $-5411.4404(19)$ &   978 [748, 1208] \\  
        TS$_{\text{I/4a}}$ &  73 & 420 & $-5409.8849$  & 280.27 & $-5409.9025(25)$         & 3448 [3115, 3781]  \\  
        PC$^-$ &  73 &           400 &  $-5409.5811$ &           229.63&  $-5409.5889(22)$ &  1094 [726, 1462] \\  
        4a &  43 &             825 &  $-5103.8866$ &           228.17&  $-5103.8879(1)$ &  1036 [972, 1100] \\ 
        \bottomrule
    \end{tabular*} 
    \label{tab:dmrg_results_mo_n2_pincer}
\end{subtable}
\end{table}
\end{savenotes}

\subsection{Classical Results Summary} 
\begin{figure}[tp]
    \centering
    \includegraphics[width=\linewidth]{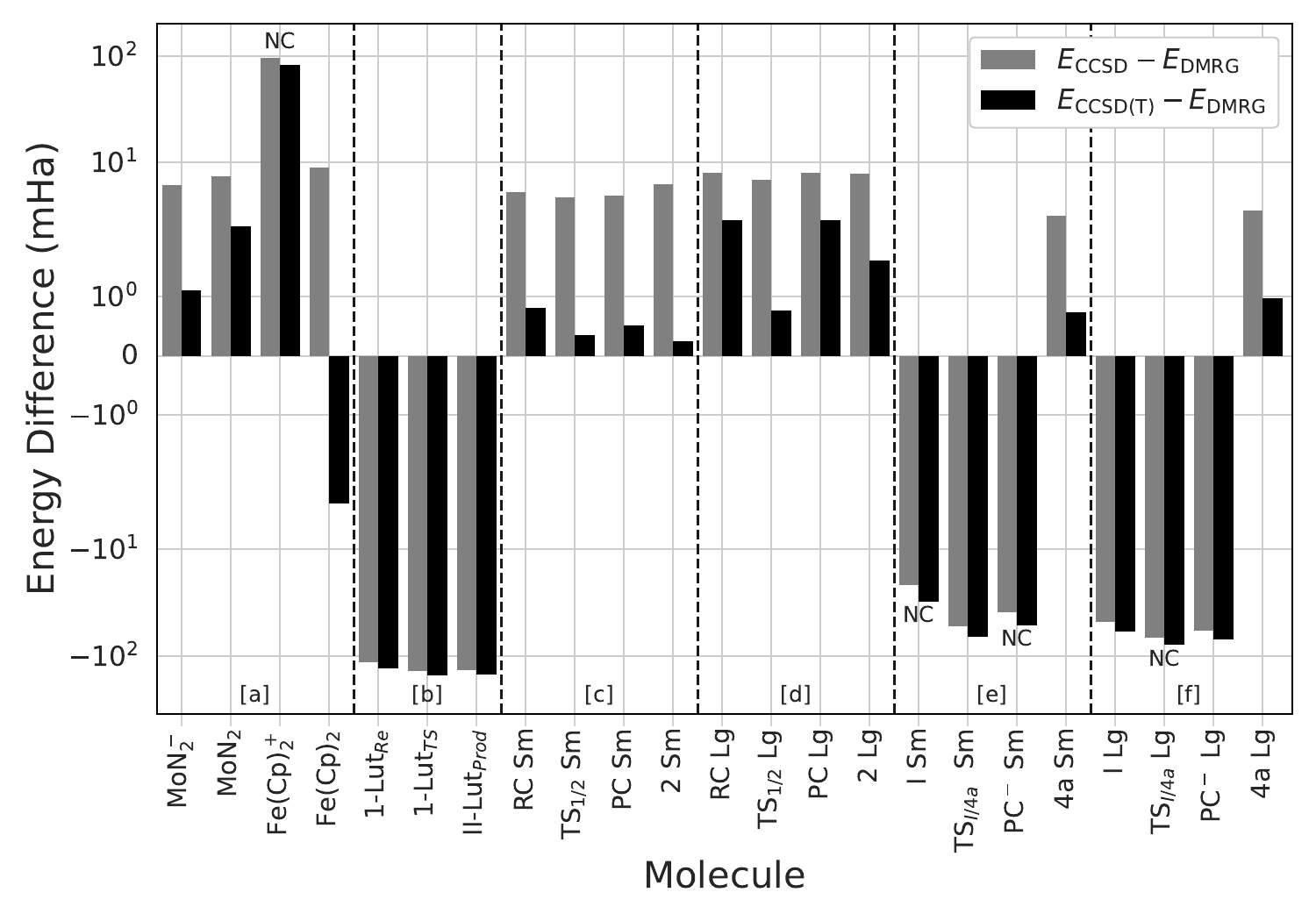}
    \caption{Differences between the DMRG extrapolated energy and the CCSD/CCSD(T) energies for the nitrogen catalyst Hamiltonians, grouped according to reaction: 
[a] Schrock;
[b] Bridged Dimolybdenum;
[c] Mo Pincer reaction, step (i) smaller active space;
[d] Mo Pincer reaction, step (i) larger active space;
[e] Mo Pincer reaction, step (ii) smaller active space;
and [f] Mo Pincer reaction, step (ii) larger active space.
``Sm'' and ``Lg'' refer to smaller and larger active spaces, respectively, and ``NC'' indicates the CCSD calculations that did not converge.}
    \label{fig:cc_dmrg_differences}
\end{figure}

To compare the CC and DMRG results, we plot the differences in the obtained energies in Figure \ref{fig:cc_dmrg_differences}. We see that no CCSD value is within 1 mHa of DMRG and only 7 CCSD(T) values are within 1 mHa of DMRG. We also see that for positive CCSD-DMRG differences, the triples correction reduces the difference to DMRG; for negative CCSD differences, the opposite is true. One notable trend is that for Hamiltonians with 52 orbitals or fewer, the differences are almost all positive; the only exception is the CCSD(T) energy for Fe(Cp)$_2$. For Hamiltonians with 55 or more orbitals, however, the differences are almost all negative. 

We can also see that all calculations where CCSD did not converge have comparatively larger absolute differences from DMRG for both CCSD and CCSD(T). However, they are not the largest differences and they also do not correspond to the largest extrapolated bond dimensions for DMRG. We note that the extrapolated bond dimension does not correlate strongly with CCSD(T)-DMRG energy differences, CCSD-DMRG energy differences, or CCSD(T) correlation energies (data not shown; absolute Pearson correlation coefficients are 0.34, 0.37 and 0.08, respectively). Taking the logarithm of the extrapolated bond dimension and the logarithm of the absolute correlation energies does improve the absolute Pearson correlation coefficient to 0.53, but this is still not a strong correlation. These results imply the difficulty of a DMRG calculation is not well-predicted by any difficulties in CCSD or CCSD(T), for the present systems.

Overall in this section, we have provided results for both CC and DMRG calculations for the nitrogen catalyst Hamiltonians discussed in Section \ref{sec:app_paper_prob_instances}.
While CC is generally a significantly faster method (the largest CC CPU time we see is 1.3 hrs, compared to typically tens to hundreds of hours for DMRG), most of the CC and DMRG results are not within 1 mHa of each other. These results, combined with the fact that DMRG is used in the literature as a reference method for large systems with large static correlation \cite{baiardiDmrg2020}, show that DMRG is the better method to benchmark against for the nitrogen catalyst Hamiltonians discussed here.

\section{Quantum Resource Estimation}
\subsection{Logical Resource Estimation}

Logical quantum resource estimates can provide insight into the feasibility of a quantum algorithm in manner that is hardware agnostic.
In the context of this work, logical quantum resource estimates are a specification of the number of non-Clifford gates, number of logical qubits, number of circuit repetitions, and tolerable hardware failure rate for each shot.
These costs are determined by models that account for various failure mode error rates as a function of algorithm parameters.
Obtaining rigorous logical quantum resource estimates presents several challenges because quantum phase estimation has many failure modes and sources of error, some of which are difficult to rigorously quantify for large systems that are near or beyond the limits of classical methods.
This section presents an algorithm performance model that captures the combined effect of important error sources and failure modes (including eigenstate projection) and reports logical estimates for the three nitrogen fixation application instances.

\subsubsection{Methods}
\label{sec:qre_logical_methods}
Quantum Phase Estimation contains a number of parameters which must be chosen so as to satisfy the performance requirements, i.e. that the algorithm should yield an estimate of the ground state energy that is within $\epsilon$ of the true ground state energy of the Hamiltonian with probability at least $1-\delta$.
This section describes the details of the quantum algorithm to be used, the free parameters in the algorithm, how these parameters affect the failure rate, and the method used to assign values to these parameters.

Although there are many variations on using quantum phase estimation for ground state energy estimation \cite{berry2000optimal,lin2022heisenberg, rendon2022effects, wang2023quantum, wang2023faster}, this work aims to choose an approach that minimizes the quantum resource requirements while still being amenable to rigorous performance analysis.
The resource estimates consider qubitized Quantum Phase Estimation (QPE) applied to the double factorized electronic structure Hamiltonian \cite{vonBurg2021}, as implemented in OpenFermion \cite{openfermion} and analyzed by Lee et al. \cite{Lee2020}.
Furthermore, to account for imperfect overlap between the initial state and the ground state, each phase estimation circuit is assumed to be repeated multiple times with the lowest obtained energy being used as the estimate of the ground state energy \cite{Berry2018}.

The algorithm parameters to be determined for a given Hamiltonian are the number of phase estimation bits $m$, the number of shots $M$, the allowable hardware failure rate $\delta'_\text{HW}$ for each shot, the number of bits of precision $\aleph$ to encode coefficients with, the number of bits of precision $\beth$ to encode rotation angles with, and the Hamiltonian truncation threshold $t$.
The sources of error considered are phase estimation spectral leakage, eigenstate projection failure, hardware failure, Hamiltonian coefficient encoding error, rotation angle encoding error, and truncation error.
Note that the last three error sources correspond to a mismatch between the Hamiltonian encoded by the block encoding and the target Hamiltonian.
Appendix \ref{sec:algorithm-perf-model} describes how these parameters influence the algorithm failure rate and the methodology used in this work for choosing algorithm parameters.

As discussed in Appendix \ref{sec:algorithm-perf-model}, the performance of QPE is dependent on the ability to efficiently prepare an initial state that has high overlap with the ground state.
This work uses the Configuration State Function (CSF) which has the highest overlap with a DMRG approximation of the ground state as the initial state.
This approach can yield a significantly higher overlap with the ground state compared to a single Slater determinant while adding only negligible quantum resources to the circuit.
These CSFs were obtained by applying block2 \cite{zhai2023block2} to the active space Hamiltonians obtained with PySCF \cite{Sun2018,Sun2020,libcint}.
The overlap between the dominant CSF and the DMRG solution to the active space Hamiltonian was used as a proxy for the overlap $\gamma$ with the ground state of the encoded Hamiltonian.
Among the Hamiltonians considered for resource estimation in this work, the CSF with the largest estimated ground-state overlap never contained more than 28 determinants.
Based on Eqs. 17 and 18 of Ref. \cite{fomichev2023initial}, the qubit and gate cost of preparing such CSFs initial state is negligible compared to the cost of the phase estimation circuit and therefore is omitted from the resource estimates presented here.

The resource estimates in this work use OpenFermion to compute the Toffoli and logical qubit counts \cite{openfermion,Lee2020} and assume an acceptable failure rate of $\overline{\delta} = 0.01$ and required accuracy of $\epsilon = 1.6$~mHa (chemical accuracy).
The analysis presented here considers only non-Clifford operations because, as discussed in Ref. \cite{Lee2020}, the costs of implementing these are expected to dwarf the costs of implementing Clifford operations for the hardware architectures considered in the physical resource estimates below.
Futhermore, note that the analysis presented here employs the Toffoli gate as the sole non-Clifford operation to enable comparison to related work \cite{Lee2020,vonBurg2021,goings2022,gidney2021factoring} that similarly analyzed resource requirements in terms of Toffoli gates and to enable physical resource estimates leveraging AutoCCZ factories to implement Toffoli gates.
To compare to prior logical resource estimates based on T gates \cite{babbush2018linear,kim2022electrolyte,reiher2017}, note that a Toffoli gate can be implemented using four T gates \cite{cody2013toffoli}.
However, it may be more meaningful to consider a Toffoli gate as roughly equivalent to two T gates in terms of physical cost because an AutoCCZ factory can be used to implement two T gates for the cost of a single Toffoli \cite{gidney2019catalyzed}.

To explore the effect of the Block-Invariant Symmetry Shift (BLISS) technique \cite{loaiza2023}, selected resource estimates presented here utilize the linear programming block-invariant symmetry shift (LPBLISS) method. LPBLISS modifies the part of the Hamiltonian spectrum that contains more or fewer electrons than the desired molecular species, with the goal of reducing the L1 norm of the Hamiltonian. The L1 norm is a key quantity that impacts the resources required by the linear combination of unitaries \cite{lowHamSim2019,childsLCU2012,loaizaSimCost2023,loaiza2023} used in our implementation of QPE. Crucially, LPBLISS performs this modification without impacting the ground state of the Hamiltonian. More details on the method can be found in Appendix \ref{app:lpbliss}, with a full description to be provided in a future manuscript \cite{2024lpbliss}.

\subsubsection{Logical Resource Estimates}
\label{sec:nitrogen_fixation_lre}
Table \ref{tab:nitrogen_fixation_lre} lists the logical resource estimates for all Hamiltonians in the nitrogen fixation instances, including the number of shots, hardware failure tolerance per shot, and number of Toffoli gates, number of logical qubits.
The estimated absolute ground-state overlap of the dominant CSF is also shown for reference.

\begin{savenotes}
\begin{table}[htp]
    \setlength\tabcolsep{0pt}
    \centering
    \caption{
    Logical resource estimates for nitrogen fixation instances. The logical resource estimate is comprised of the number of orbitals $N_o$, the number of shots $M$, hardware failure tolerance per shot $\delta'_{\text{HW}}$, number of Toffoli gates per shot $N_\text{Toffoli}$, number of logical qubits $N_\text{q}$. The absolute overlap $\left|\gamma\right|$ between the dominant CSF and the ground state (estimated with DMRG) is also shown for reference.} \label{tab:nitrogen_fixation_lre}

    \begin{subtable}{\textwidth}
    \caption{Logical resource estimates for Schrock Catalyst (Section \ref{sec:schrock_mo_n2_instance})}    \label{tab:mo_n2_LRE}
    \begin{tabular*}{\textwidth}{@{\extracolsep{\fill}}lcccccc}
        \toprule
        Molecule ID & $N_o$ & $M$ & $\delta'_{\text{HW}}$ & $N_\text{Toffoli}$ & $N_\text{q}$ & $\left|\gamma\right|$\\ \hline
        
        MoN$_2^-$        & 33    & 3 & $3.33 \times 10^{-4}$ & $7.7 \times 10^{10}$ & 1914 & 0.97 \\ 
        MoN$_2$          & 33  & 3 & $3.33 \times 10^{-4}$ & $7.7 \times 10^{10}$ & 1921 & 0.96 \\ 
        Fe(Cp)$_2^+$     & 46     & 8 & $1.25 \times 10^{-4}$ & $4.1 \times 10^{11}$ & 2697 & 0.78 \\ 
        Fe(Cp)$_2$       & 46    & 3 & $3.33 \times 10^{-4}$ & $2.5 \times 10^{11}$ & 2695 & 0.97 \\ 
        \bottomrule
    \end{tabular*}
    \end{subtable}
    
    \vspace*{5mm}
    
    \begin{subtable}{\textwidth}
    \caption{Logical resource estimates for Bridged Dimolybdenum  (Section \ref{sec:mo2_n2_instance})}    \label{tab:mo2_n2_LRE}
    \begin{tabular*}{\textwidth}{@{\extracolsep{\fill}}lcccccc}
        \toprule
        Molecule ID & $N_o$ & $M$ & $\delta'_{\text{HW}}$ & $N_\text{Toffoli}$ & $N_\text{q}$ & $\left|\gamma\right|$\\ \hline
         1-Lut$_{Re}$     & 69    & 5 & $2.00 \times 10^{-4}$ & $1.1 \times 10^{12}$ & 7816 & 0.88 \\ 
         1-Lut$_{TS}$     & 70    & 5 & $2.00 \times 10^{-4}$ & $1.1 \times 10^{12}$ & 7925 & 0.88 \\ 
         II-Lut$_{Prod}$  & 70    & 5 & $2.00 \times 10^{-4}$ & $1.1 \times 10^{12}$ & 7925 & 0.88 \\ 
        \bottomrule
    \end{tabular*}
    \end{subtable}
    
    \vspace*{5mm}

    \begin{subtable}{\textwidth}
    \caption{Logical resource estimates for Mo Pincer reactions (Section \ref{sec:mo_n2_pincer_instance})} \label{tab:mo_n2_pincer_LRE}
    \begin{tabular*}{\textwidth}{@{\extracolsep{\fill}}lcccccc}
        \toprule
        Molecule ID & $N_o$ & $M$ & $\delta'_{\text{HW}}$ & $N_\text{Toffoli}$ & $N_\text{q}$ & $\left|\gamma\right|$\\ \hline
        \multicolumn{6}{c}{step (i) smaller active space} \\ \hline
        RC               & 32 & 3 & $3.33 \times 10^{-4}$ & $7.7 \times 10^{10}$ & 1062 & 0.96 \\ 
        TS$_{1/2}$       & 27 & 3 & $3.33 \times 10^{-4}$ & $5.4 \times 10^{10}$ & 928 & 0.97 \\ 
        PC               & 32 & 3 & $3.33 \times 10^{-4}$ & $9.8 \times 10^{10}$ & 1865 & 0.96 \\ 
        2                & 33 & 3 & $3.33 \times 10^{-4}$ & $1.1 \times 10^{11}$ & 1922 & 0.96 \\ 
        \hline
        \multicolumn{6}{c}{step (i) larger active space} \\ \hline
        RC               & 51 & 2 & $5.00 \times 10^{-4}$ & $1.9 \times 10^{11}$ & 1639 & 0.99 \\ 
        TS$_{1/2}$       & 51 & 4 & $2.50 \times 10^{-4}$ & $3.6 \times 10^{11}$ & 2972 & 0.94 \\ 
        PC               & 51 & 4 & $2.50 \times 10^{-4}$ & $4.1 \times 10^{11}$ & 2964 & 0.98 \\ 
        2                & 52 & 5 & $2.00 \times 10^{-4}$ & $3.4 \times 10^{11}$ & 1663 & 0.88 \\ 
        \hline
        \multicolumn{6}{c}{step (ii) smaller active space} \\ \hline
        I                & 56 & 5 & $2.00 \times 10^{-4}$ & $5.3 \times 10^{11}$ & 3352 & 0.87 \\ 
        TS$_{\text{I/4a}}$      & 56 & 9 & $1.11 \times 10^{-4}$ & $7.1 \times 10^{11}$ & 3244 & 0.74 \\ 
        PC$^-$           & 55 & 4 & $2.50 \times 10^{-4}$ & $4.1 \times 10^{11}$ & 3187 & 0.94 \\ 
        4a               & 24 & 7 & $1.43 \times 10^{-4}$ & $6.3 \times 10^{10}$ & 847 & 0.82 \\ 
        \hline
        \multicolumn{6}{c}{step (ii) larger active space} \\ \hline
        I                & 75 & 5 & $2.00 \times 10^{-4}$ & $1.4 \times 10^{12}$ & 8478 & 0.86 \\ 
        TS$_{\text{I/4a}}$      & 73 & 6 & $1.67 \times 10^{-4}$ & $1.4 \times 10^{12}$ & 8258 & 0.85 \\
        PC$^-$           & 73 & 5 & $2.00 \times 10^{-4}$ & $9.9 \times 10^{11}$ & 4313 & 0.87 \\         
        4a               & 43 & 6 & $1.67 \times 10^{-4}$ & $2.6 \times 10^{11}$ & 2532 & 0.85 \\ 
        \bottomrule
    \end{tabular*}
    \end{subtable}
\end{table}
\end{savenotes}

This data shows that the number of active-space orbitals plays has a strong influence on the number of logical qubits and Toffoli gates.
For example, consider the Hamiltonians with the most active-space orbitals, Bridged Dimolydenum and \ch{Mo} Pincer step (ii) in the large active space.
These Hamiltonians all have $\sim 70$ active-space orbitals and require $\sim 10^12$ Toffoli gates.
The other Hamiltonians, in contrast, have significantly less that 70 active-space orbitals and require significantly fewer Toffoli gates.

Table \ref{tab:nitrogen_fixation_lre} also illustrates the effect of ground-state overlap $\left|\gamma\right|$ on the the number of shots $M$ increases and the the acceptable hardware failure rate per shot $\delta'_\text{HW}$.
In particular, as the ground-state overlap decreases, the number of shots required increases and acceptable hardware failure rate per shot decreases, as expected from Eqs. \ref{eq:num-shots} and \ref{eqn:hw-failure-rate}.
For example, the Hamiltonian with the lowest absolute ground-state overlap, TS$_{\text{I/4a}}$ with a 56 orbital active space ($\left|\gamma\right| = 0.74$), has the largest number of shots ($M=9$) and the lowest acceptable hardware failure rate per shot ($\delta'_\text{HW}=1.11 \times 10^{-4}$).

Table \ref{tab:nitrogen_fixation_lre_lpbliss} shows logical resource estimates for the LPBLISS-treated Schrock and Bridged Dimolybdenum catalyst Hamiltonians. For the Schrock catalyst, the number of Toffoli gates dropped by a factor of about two in comparison to the original Hamiltonians (Table \ref{tab:nitrogen_fixation_lre}). The number of logical qubits also dropped by about 4\%. For the Bridged Dimolybdenum catalyst, both the number of Toffoli gates and the number of logical qubits dropped by a factor of about two.
These results suggests that methods such as LPBLISS can significantly reduce the logical resource requirements of high-utility GSEE instances.


\begin{savenotes}
\begin{table}[htp]
    \setlength\tabcolsep{0pt}
    \centering
    \caption{
    Logical resource estimates for LPBLISS treated nitrogen fixation instances Hamiltonians. The logical resource estimate is comprised of the number of orbitals $N_o$, the number of shots $M$, hardware failure tolerance per shot $\delta'_{\text{HW}}$, number of Toffoli gates per shot $N_\text{Toffoli}$, number of logical qubits $N_\text{q}$. The absolute overlap $\left|\gamma\right|$ between the dominant CSF and the ground state (estimated with DMRG) is also shown for reference.} \label{tab:nitrogen_fixation_lre_lpbliss}

    \begin{subtable}{\textwidth}
    \caption{Logical resource estimates for LPBLISS treated Schrock Catalyst (Section \ref{sec:schrock_mo_n2_instance})}    \label{tab:mo_n2_LRE_BLISS}
    \begin{tabular*}{\textwidth}{@{\extracolsep{\fill}}lcccccc}
        \toprule
        Molecule ID & $N_o$ & $M$ & $\delta'_{\text{HW}}$ & $N_\text{Toffoli}$ & $N_\text{q}$ & $\left|\gamma\right|$\\ \hline
        MoN$_2^-$        & 33 & 3 & $3.33 \times 10^{-4}$ & $3.1 \times 10^{10}$ & 1849 & 0.97 \\ 
        MoN$_2$          & 33 & 3 & $3.33 \times 10^{-4}$ & $3.2 \times 10^{10}$ & 1849 & 0.97 \\ 
        Fe(Cp)$_2^+$     & 46 & 8 & $1.25 \times 10^{-4}$ & $2.0 \times 10^{11}$ & 2602 & 0.78 \\ 
        Fe(Cp)$_2$       & 46 & 3 & $3.33 \times 10^{-4}$ & $1.4 \times 10^{11}$ & 2601 & 0.97 \\ 
        \bottomrule
    \end{tabular*}
    \end{subtable}
    
    \vspace*{5mm}
    
    \begin{subtable}{\textwidth}
    \caption{Logical resource estimates for LPBLISS treated Bridged Dimolybdenum (Section \ref{sec:mo2_n2_instance})}    \label{tab:mo2_n2_LRE_BLISS}
    \begin{tabular*}{\textwidth}{@{\extracolsep{\fill}}lcccccc}
        \toprule
        Molecule ID & $N_o$ & $M$ & $\delta'_{\text{HW}}$ & $N_\text{Toffoli}$ & $N_\text{q}$ & $\left|\gamma\right|$\\ \hline
        1-Lut$_{Re}$     & 69 & 5 & $2.00 \times 10^{-4}$ & $5.9 \times 10^{11}$ & 4081 & 0.88 \\ 
        1-Lut$_{TS}$     & 70 & 5 & $2.00 \times 10^{-4}$ & $6.0 \times 10^{11}$ & 4137 & 0.88 \\
        II-Lut$_{Prod}$  & 70 & 5 & $2.00 \times 10^{-4}$ & $5.2 \times 10^{11}$ & 4136 & 0.88 \\ 
        \bottomrule
    \end{tabular*}
    \end{subtable}
\end{table}
\end{savenotes}
\subsection{Physical Resource Estimation}
\label{subsec:phys_qre}
The physical resource estimates presented in this section were made using existing models of logical architecture compilation, magic state distillation, surface code performance, and the physical architecture capabilities.
The input data from the logical resource estimates that is used to generate the physical resource estimates is simply the number of logical qubits, the number of Toffoli gates, the number of parallel magic state factories, and the per-circuit tolerable failure rate.

We adopt the physical resource estimation methodology used in \cite{goings2022} and overview this method in Section \ref{app:logical_architecture_model}.
Here, the primitive logical operations are
assumed to be nearest-neighbor lattice surgery operations facilitated by the surface code \cite{horsman2012surface} and magic state preparation facilitated by Litinski factories \cite{litinski2019magic}, with each Toffoli gate requiring four distillations \cite{jones2013toffoli}.
The physical architecture supporting these operations is a two-dimensional grid of superconducting qubits assumed to have surface code cycle times of 100~ns and a physical error rate of 0.001.
The physical resource estimates were performed using Bench-Q \cite{BenchQ}, which in turn relies on physical costing features provided by OpenFermion \cite{openfermion}.

\begin{figure}[ht]
    \centering
    \includegraphics[width=0.6\linewidth]{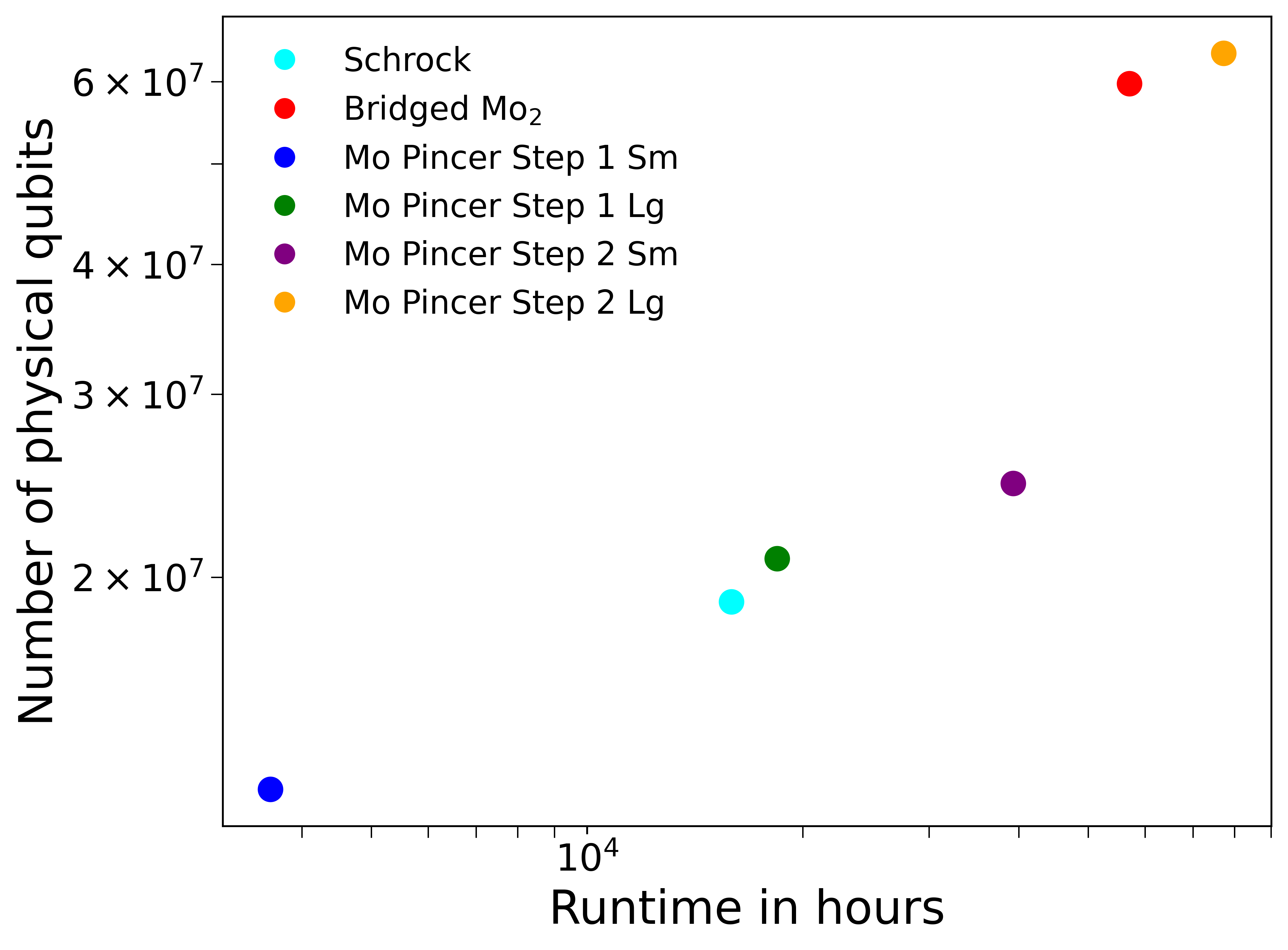}
    \caption{Physical resource estimates for the nitrogen fixation reactions described in Section \ref{sec:app_paper_prob_instances}. Aggregated resource estimates for reactions including Bridged Dimolybdenum, Schrock catalyst and Mo Pincer step (i) and (ii) are displayed separately as independent dots. The horizontal axis represents the total runtime, accounting for all Hamiltonians in the reaction and all circuit repetitions, and the vertical axis of the plot represents the maximum number of physical qubits required by any Hamiltonian in that reaction. `Sm' and `Lg' represent different active space selections.}
    \label{fig:nitrogen_fixation_pre}
\end{figure}

Figure \ref{fig:nitrogen_fixation_pre} illustrates the runtime and qubit requirements for the serial execution of all shots required for the nitrogen fixation instances. The estimated number of physical qubits represents maximum number required by each reaction, and the total runtime represents the summation of runtimes for all Hamiltonians in each individual reaction. It is important to note that this analysis assumes no parallelization of shots; each is processed sequentially. The physical estimates are based on the preliminary logical resources detailed in Table \ref{tab:nitrogen_fixation_lre}.
Table \ref{tab:nitrogen_fixation_pre} shows the runtime and number of physical qubits required for each Hamiltonian.
Figure \ref{fig:nitrogen_fixation_pre}  and Table \ref{tab:nitrogen_fixation_pre} illustrate the importance of the number of active-space orbitals.
For example, the instances containing Hamiltonians with the greatest number of active-space orbtials, Bridged Dimolydenum and Mo Pincer (ii) in the large active space, also require the greatest number of physical qubits and runtime.

\begin{savenotes}
\begin{table}[htp]
    \setlength\tabcolsep{0pt}
    \centering
    \caption{
    Physical resource estimates for nitrogen fixation instances. The physical resource estimate is comprised of the number of orbitals $N_o$, the number of shots $M$, the number of physical qubit $N_\text{phys}$ and the total runtime in hours $T_\text{hr}$.} \label{tab:nitrogen_fixation_pre}

    \begin{subtable}{\textwidth}
    \caption{Physical resource estimates for Schrock Catalyst (Section \ref{sec:schrock_mo_n2_instance})}    \label{tab:mo_n2_PRE}
    \begin{tabular*}{\textwidth}{@{\extracolsep{\fill}}lcccc}
        \toprule
        Molecule ID & $N_o$ & $M$ & $T_\text{hr}$ & $N_\text{phys}$\\ \hline
        MoN$_2^-$        & 33 & 3 & $8.18 \times 10^{2}$ & $1.24 \times 10^{6}$ \\
        MoN$_2$          & 33 & 3 & $8.20 \times 10^{2}$ & $1.25 \times 10^{6}$ \\
        Fe(Cp)$_2^+$     & 46 & 8 & $1.16 \times 10^{4}$ & $1.89 \times 10^{6}$ \\
        Fe(Cp)$_2$       & 46 & 3 & $2.67 \times 10^{3}$ & $1.89 \times 10^{6}$ \\
        \bottomrule
    \end{tabular*}
    \end{subtable}
    
    \vspace*{5mm}
    
    \begin{subtable}{\textwidth}
    \caption{Physical resource estimates for Bridged Dimolybdenum (Section \ref{sec:mo2_n2_instance})}    \label{tab:mo2_n2_PRE}
    \begin{tabular*}{\textwidth}{@{\extracolsep{\fill}}lcccc}
        \toprule
        Molecule ID & $N_o$ & $M$ & $T_\text{hr}$ & $N_\text{phys}$ \\ \hline
        1-Lut$_{Re}$     & 69 & 5 & $1.88 \times 10^{4}$ & $5.89 \times 10^{7}$ \\
        1-Lut$_{TS}$     & 70 & 5 & $1.91 \times 10^{4}$ & $5.97 \times 10^{7}$ \\ 
        II-Lut$_{Prod}$  & 70 & 5 & $1.92 \times 10^{4}$ & $5.97 \times 10^{7}$ \\
        \bottomrule
    \end{tabular*}
    \end{subtable}
    
    \vspace*{5mm}

    \begin{subtable}{\textwidth}
    \caption{Physical resource estimates for Mo Pincer reactions (Section \ref{sec:mo_n2_pincer_instance})} \label{tab:mo_n2_pincer_PRE}
    \begin{tabular*}{\textwidth}{@{\extracolsep{\fill}}lcccc}
        \toprule
        Molecule ID & $N_o$ & $M$ & $T_\text{hr}$ & $N_\text{phys}$ \\ \hline
        \multicolumn{3}{c}{step (i) smaller active space} \\ \hline
        RC               & 32 & 3 & $8.18 \times 10^{2}$ & $7.04 \times 10^{5}$ \\
        TS$_{1/2}$       & 27 & 3 & $5.77 \times 10^{2}$ & $6.18 \times 10^{5}$ \\
        PC               & 32 & 3 & $1.05 \times 10^{3}$ & $1.21 \times 10^{6}$ \\
        2                & 33 & 3 & $1.17 \times 10^{3}$ & $1.25 \times 10^{6}$ \\

        \hline
        \multicolumn{3}{c}{step (i) larger active space} \\ \hline
        RC               & 51 & 2 & $1.33 \times 10^{3}$ & $1.07 \times 10^{6}$ \\
        TS$_{1/2}$       & 51 & 4 & $5.18 \times 10^{3}$ & $2.08 \times 10^{6}$ \\
        PC               & 51 & 4 & $5.84 \times 10^{3}$ & $2.08 \times 10^{6}$ \\
        2                & 52 & 5 & $6.06 \times 10^{3}$ & $1.18 \times 10^{6}$ \\

        \hline
        \multicolumn{3}{c}{step (ii) smaller active space} \\ \hline
        I                & 56 & 5 & $9.35 \times 10^{3}$ & $2.35 \times 10^{6}$ \\
        TS$_{\text{I/4a}}$      & 56 & 9 & $2.26 \times 10^{4}$ & $2.46 \times 10^{6}$ \\
        PC$^-$           & 55 & 4 & $5.76 \times 10^{3}$ & $2.23 \times 10^{6}$ \\
        4a               & 24 & 7 & $1.56 \times 10^{3}$ & $5.67 \times 10^{5}$ \\
        \hline
        \multicolumn{3}{c}{step (ii) larger active space} \\ \hline
        I                & 75 & 5 & $2.47 \times 10^{4}$ & $6.39 \times 10^{6}$ \\
        TS$_{\text{I/4a}}$      & 73 & 6 & $3.08 \times 10^{4}$ & $6.23 \times 10^{6}$ \\
        PC$^-$           & 73 & 5 & $1.62 \times 10^{4}$ & $3.26 \times 10^{6}$ \\
        4a               & 43 & 6 & $5.56 \times 10^{3}$ & $1.78 \times 10^{6}$ \\
        \bottomrule
    \end{tabular*}
    \end{subtable}
\end{table}
\end{savenotes}

\begin{figure}[htp]
    \centering
    \includegraphics[width=0.6\linewidth]{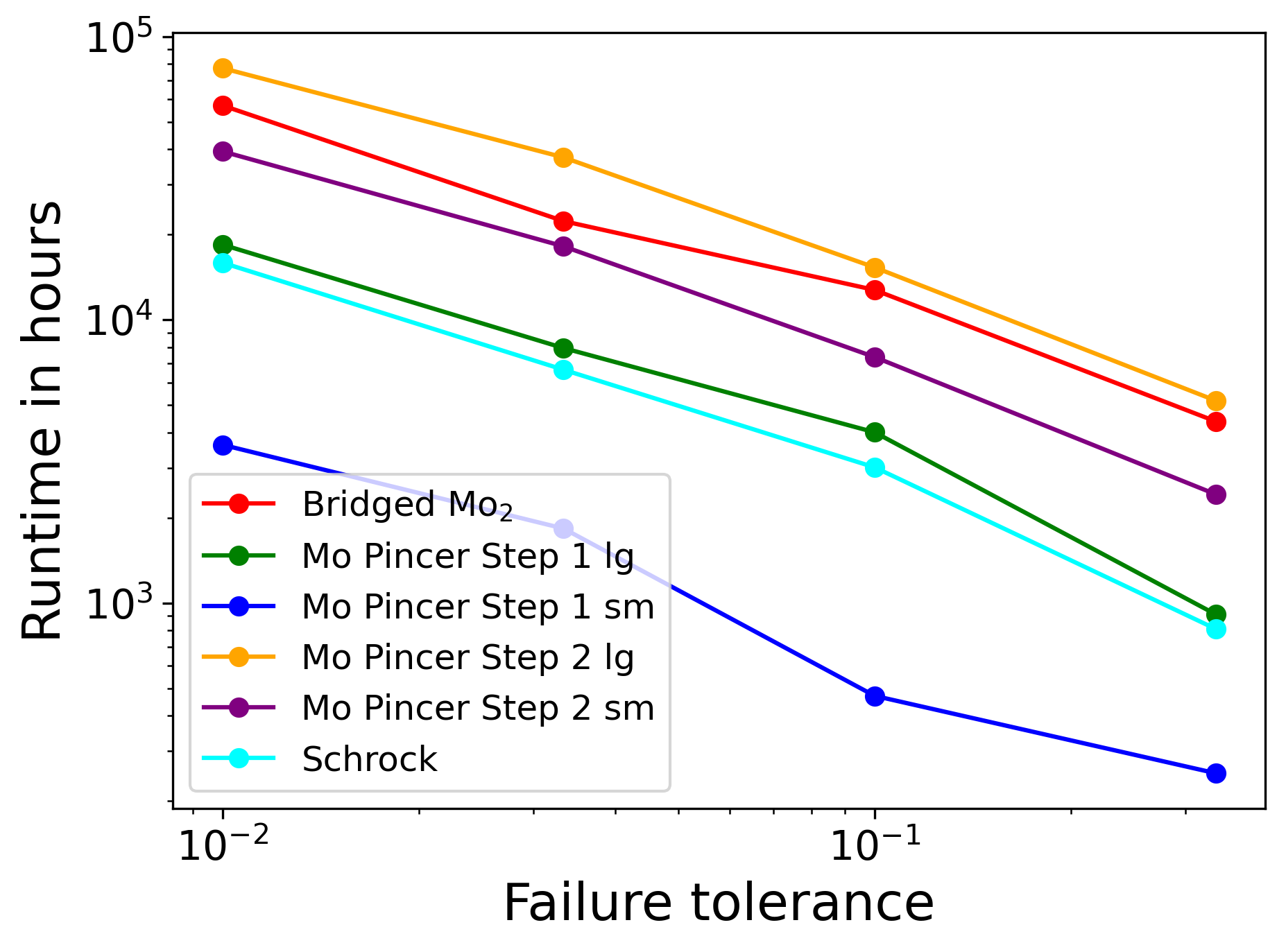}
    \caption{Estimated runtime for the nitrogen fixation reactions described in Section \ref{sec:app_paper_prob_instances} versus failure rate tolerance. Aggregated resource estimates for reactions including Bridged Dimolybdenum, Schrock catalyst and Mo Pincer step (i) and (ii) are displayed separately as independent dots. `sm' and `lg' represent different active space selections.}
    \label{fig:runtime-vs-failure-rate}
\end{figure}

Fig. \ref{fig:runtime-vs-failure-rate} shows how the runtime decreases as the allowable failure rate $\overline{\delta}$ is increased.
Increasing the allowable failure rate from 0.01 to 0.33 decreases the runtime by approximately a factor of 10.

\begin{figure}[htp]
    \centering
    \includegraphics[width=0.6\linewidth]{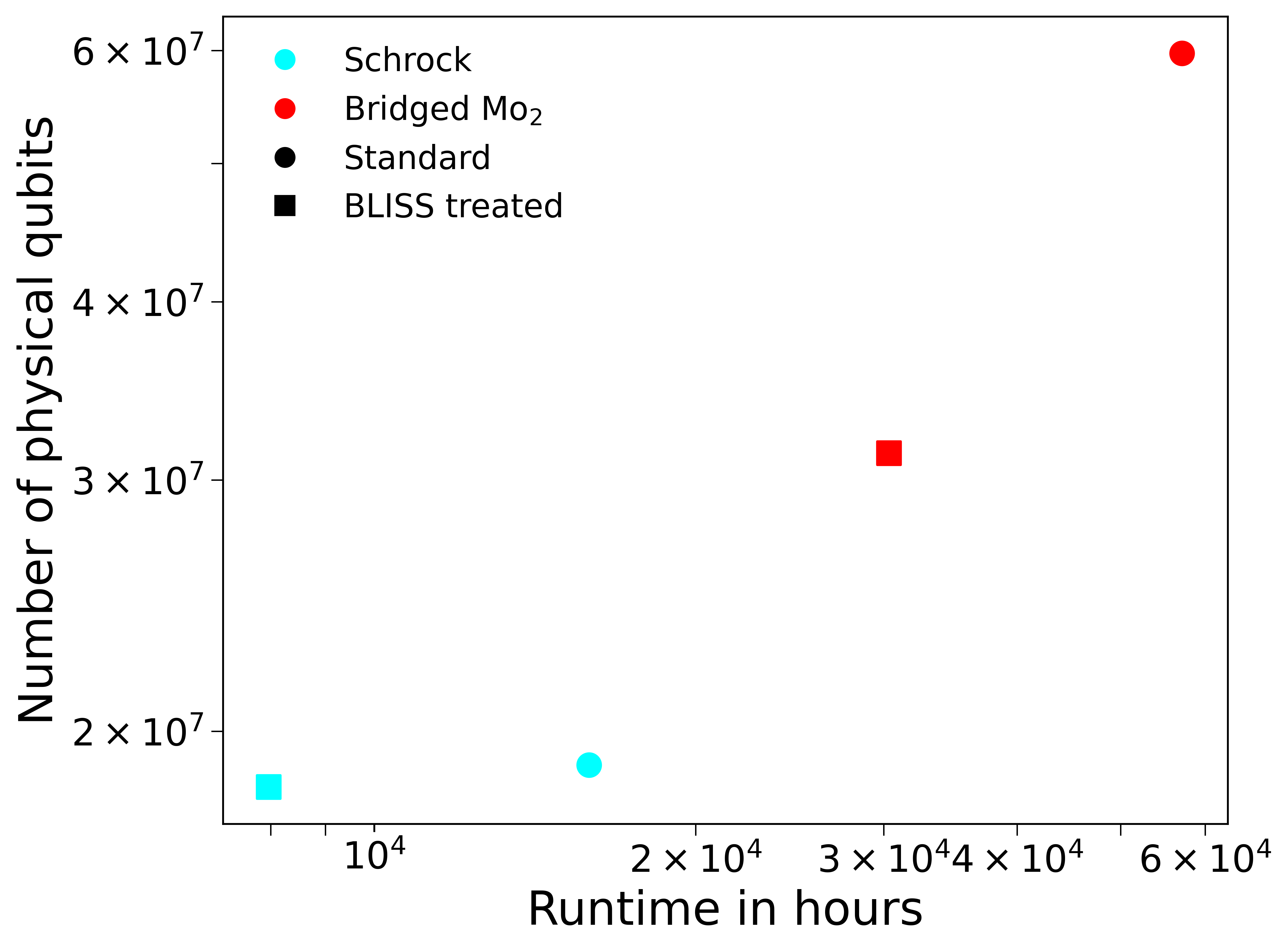}
    \caption{Physical resource estimates for LPBLISS treated nitrogen fixation reactions described in Section \ref{sec:app_paper_prob_instances}. Aggregated resource estimates for reactions including Bridged Dimolybdenum, Schrock catalyst and Mo Pincer step (i) and (ii) are displayed separately as independent dots. The horizontal axis represents the total runtime, accounting for all Hamiltonians in the reaction and all circuit repetitions, and the vertical axis of the plot represents the maximum number of physical qubits required by any Hamiltonian in that reaction. `Sm' and `Lg' represent different active space selections.}
    \label{fig:nitrogen_fixation_pre_LPBLISS}
\end{figure}

\begin{savenotes}
\begin{table}[htp]
    \setlength\tabcolsep{0pt}
    \centering
    \caption{
    Physical resource estimates for LPBLISS treated 
    nitrogen fixation Hamiltonians. The physical resource estimate is comprised of the number of orbitals $N_o$, the number of shots $M$, the number of physical qubit $N_\text{phys}$ and the total runtime in hours $T_\text{hr}$.} \label{tab:nitrogen_fixation_pre_LPBLISS}

    \begin{subtable}{\textwidth}
    \caption{Physical resource estimates for LPBLISS treated Schrock Catalyst (Section \ref{sec:schrock_mo_n2_instance})}    \label{tab:mo_n2_PRE_LPBLISS}
    \begin{tabular*}{\textwidth}{@{\extracolsep{\fill}}lcccc}
        \toprule
        Molecule ID & $N_o$ & $M$ & $T_\text{hr}$ & $N_\text{phys}$\\ \hline
        MoN$_2^-$        & 33 & 3 & $3.30 \times 10^{2}$ & $1.20 \times 10^{6}$ \\
        MoN$_2$          & 33 & 3 & $3.45 \times 10^{2}$ & $1.20 \times 10^{6}$ \\
        Fe(Cp)$_2^+$     & 46 & 8 & $5.81 \times 10^{3}$ & $1.83 \times 10^{6}$ \\
        Fe(Cp)$_2$       & 46 & 3 & $1.48 \times 10^{3}$ & $1.68 \times 10^{6}$ \\
        \bottomrule
    \end{tabular*}
    \end{subtable}
    
    \vspace*{5mm}
    
    \begin{subtable}{\textwidth}
    \caption{Physical resource estimates for LPBLISS treated Bridged Dimolybdenum (Section \ref{sec:mo2_n2_instance})}    \label{tab:mo2_n2_PRE_LPBLISS}
    \begin{tabular*}{\textwidth}{@{\extracolsep{\fill}}lcccc}
        \toprule
        Molecule ID & $N_o$ & $M$ & $T_\text{hr}$ & $N_\text{phys}$\\ \hline
        1-Lut$_{Re}$     & 69 & 5 & $1.04 \times 10^{4}$ & $3.09 \times 10^{6}$ \\
        1-Lut$_{TS}$     & 70 & 5 & $1.07 \times 10^{4}$ & $3.13 \times 10^{6}$ \\
        II-Lut$_{Prod}$  & 70 & 5 & $9.25 \times 10^{3}$ & $2.89 \times 10^{6}$ \\
        \bottomrule
    \end{tabular*}
    \end{subtable} 
\end{table}
\end{savenotes}

Figure \ref{fig:nitrogen_fixation_pre_LPBLISS} and Table \ref{tab:nitrogen_fixation_pre_LPBLISS} provides physical resource estimates for the LPBLISS-treated Schrock and Bridged Dimolybdenum catalyst Hamiltonians. For the Schrock catalyst, LPBLISS lowers the runtime by a factor of about two and the number of physical qubits by about 6\%, with a mean spacetime volume reduction of 2.28. For the Bridged Dimolybdenum catalyst, the reductions are more significant. The runtime dropped by a factor of about 2, while the number of physical qubits dropped by a factor of about 20, leading to a mean space-time volume drop of about 37.
While we caution against over-generalization, these results for the LPBLISS-treated Hamiltonians provide a promising example of the potential for future reductions in the amount of quantum resources required for QPE.

\section{Utility Estimation}
This work aims to elucidate the future utility of full-scale fault-tolerant quantum computers in chemistry, with a particular focus on the field of nitrogen fixation catalyst development. This area was chosen due to its significant implications for global agriculture and environmental sustainability. Currently, ammonia production, a key process in nitrogen fixation, is both energy-intensive and a substantial contributor to global carbon emissions due to its reliance on the Haber-Bosch process. By employing quantum computers to tackle complex Hamiltonians associated with nitrogen fixation catalysts, we anticipate not only enhancements in catalyst design that could lead to reductions in both costs and environmental impact. This would lead to significant advancements in food security globally. This section critically evaluates the economic viability and practical utility of quantum computational solutions in contrast to existing classical methods. We look at the economic impact of ammonia, the cost of recent catalysis research, and finally, the cost of classical replacements.

\subsection{Economic Impact of Ammonia Production}
Nitrogen fixation is a critical chemical process that transforms atmospheric nitrogen (N$_2$) into a variety of essential molecular products, pivotal in modern society. Ammonia (NH$_3$), produced through this process, serves as a foundational component for fertilizers, pharmaceuticals, and explosives, highlighting its substantial economic and industrial importance.

Ammonia production is a colossal industry, deeply reliant on fossil fuels, especially natural gas. Over recent years, ammonia prices in the United States have shown considerable volatility, affected by fluctuations in natural gas prices—from lows of around \$200 per metric ton in 2019~\cite{2019_ammonia} to highs surpassing \$1,000 per metric ton in 2022~\cite{2022_ammonia}. Annually, around 170 million metric tonnes of ammonia are produced, valuing the lower end of the market at approximately \$34 billion.

The production of ammonia is notably energy-intensive and significantly contributes to global carbon emissions. The Haber-Bosch process, the predominant method for industrial ammonia synthesis, alone is responsible for approximately 1.8\% of global carbon emissions. This substantial environmental footprint underscores an urgent need for more sustainable and less carbon-intensive alternatives.

Breaking the inert triple bond in atmospheric nitrogen, which requires about 228.6 kcal/mol (or roughly 115,000 K), poses a significant scientific challenge. The Haber-Bosch process, despite its widespread use, demands extreme operational conditions—pressures ranging from 250 to 350 bar and temperatures between 450 to 550$^\circ$C, with a heterogeneous iron catalyst, and achieves only a 10-20\% yield per cycle. These severe requirements amplify the environmental and economic costs of ammonia production.

Consequently, there is a compelling need for innovations that are less damaging to the environment and less reliant on high energy inputs. This necessity has spurred research aimed at optimizing and decarbonizing ammonia production. Studies include exploring biological nitrogen fixation mechanisms that operate under milder conditions and developing synthetic nitrogen fixation catalysts. Among these, molybdenum-based (Mo-based) catalysts are one of the most thoroughly investigated systems, presenting promising pathways for future technological advancements. These efforts not only aim to enhance the efficiency and sustainability of ammonia production but also strive to align industrial practices with environmental conservation goals.

\subsection{Funding and Research Costs in Catalysis}
In the realm of state-of-the-art research in catalysis, the pricing of computational studies reflects their significant value. A pertinent example is the Liquid Sunlight Alliance (LiSA)\cite{LiquidSunlightAlliance2024}, which has been supported by a substantial \$25 million grant from the Department of Energy over five years, starting in late 2020. Since its inception, LiSA has produced 141 papers, with about one-third dedicated solely to computational studies. Based on the project's trajectory, it is expected to yield approximately 250 papers by its conclusion, averaging a funding rate of \$100,000 per paper.

These studies often involve multiple reactions within a single paper, typically supported by various funding sources—commonly three to six. This diversity in funding not only underscores the broad interest and potential impact of the research but also helps distribute the financial risk associated with these ambitious projects. The complexity and the multi-reaction scope of these papers imply that the computational costs are substantial. Given the depth and breadth of the analysis required, a significant portion of the funding is presumably allocated to computational resources.

Thus, while each paper might be valued at more than \$100,000 due to the multiple reactions it explores, a practical estimate of \$100,000 per reaction effectively captures the computational expenditure involved. For our instances, we can expect to cap the value in the range of \$100,000 to \$200,000, depending on the number of reactions studied. 

\subsection{Assessing the Value of Computational Approaches}
A final method to assess the value of computational approaches is to compare the cost to that of classical computations.
As discussed in Section \ref{sec:dmrg}, \verb|Block2| is an efficient and highly scalable implementation of the Density Matrix Renormalization Group (DMRG) for quantum chemistry based on Matrix Product Operator formalism~\cite{zhai2023block2}.
The computational aspects of the method and the results of DMRG calculations on the nitrogen fixation instances are presented in Section \ref{sec:dmrg}.
The key parameter of the wavefunction approximation used by DMRG is the corresponding bond dimension, \(D\).
We can estimate the expected cost to achieve chemical accuracy by extrapolating the results obtained for low values of \(D\) as a function of the (approximate) ground state energy error.
Once the required bond dimension is known, we can use the computational scaling of DMRG to estimate the runtime, which is cubic in \(D\) in the leading order, \(T_{\text{DMRG}} = \mathcal{O}(D^3)\)\cite{Wouters2014}.

\begin{figure}[hpt]
    \centering
    \includegraphics[width=0.6\textwidth]{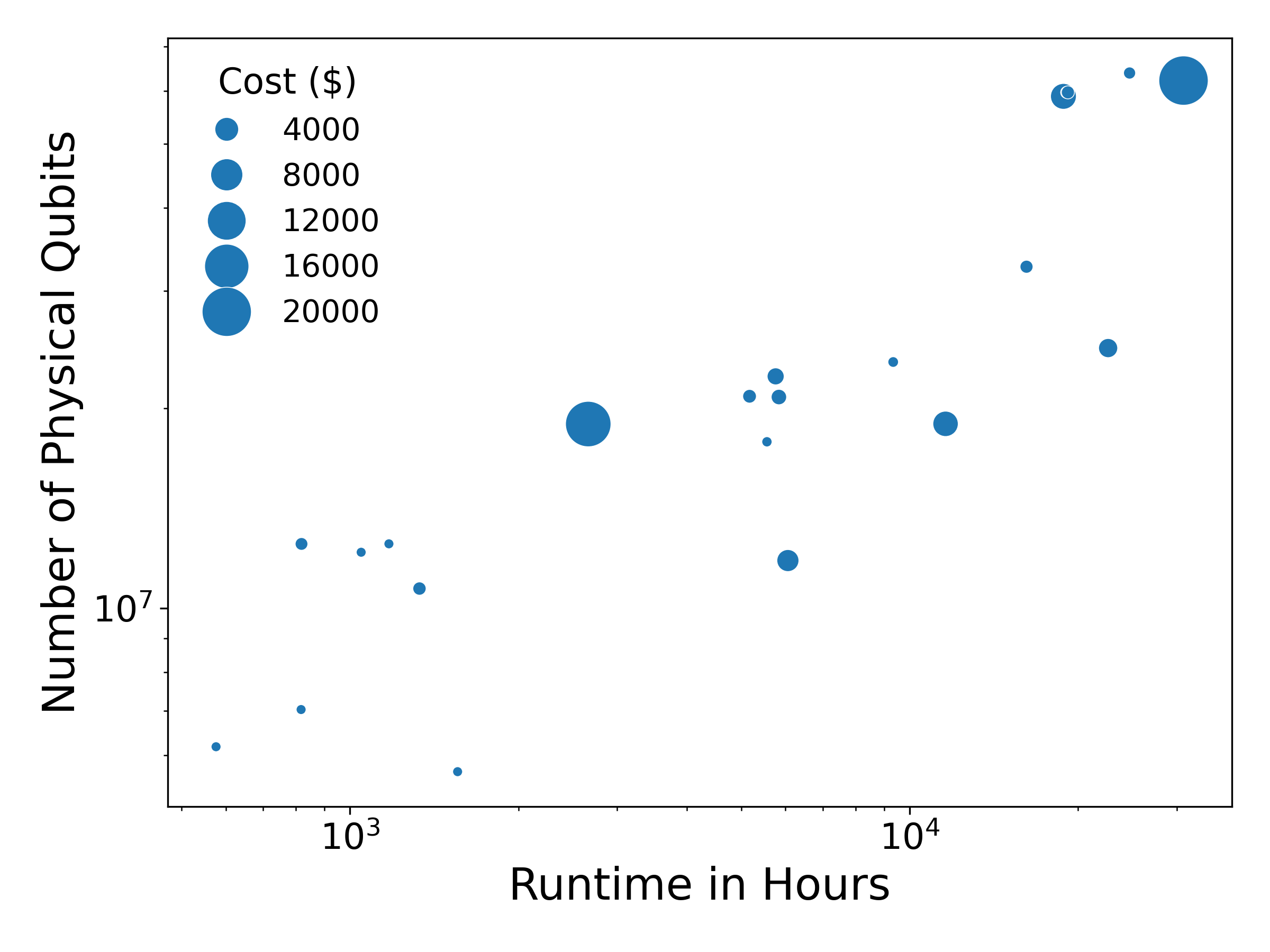}
    \caption{Runtime vs. Number of Qubits for High-Utility Instances. 
    This plot compares the classical computational costs with our standard QPE estimates, highlighting the value proposition of quantum approaches for instances involving more than 50 spatial orbitals.}
    \label{fig:runtime_vs_qubits_value}
\end{figure}

We have found that that for high-utility instances, specifically those involving more than 50 spatial orbitals, the average extrapolated CPU-hours are approximately 65,000.
At a cost of \$0.04 per CPU-hour\cite{amazon_ec2_pricing}, this translates to an average expenditure of \$2,800 per instance. 
We can use this cost comparison to highlight the economic value of these computations when benchmarked against classical methods.
Figure \ref{fig:runtime_vs_qubits_value} illustrates the runtime versus the number of qubits, comparing classical replacement costs to our vanilla Quantum Phase Estimation (QPE) estimates.
This high costs highlights the large resources required to converge to chemical accuracy (1 kcal/mol).
This comparison underscores the potential cost-effectiveness and efficiency of quantum computations in handling complex chemical reactions.

\section{Conclusion}

In this work, we have evaluated the potential of full-scale fault-tolerant quantum computers to address complex problems in nitrogen fixation catalyst development.
The results show that under a conservative model of algorithm performance, a moderately optimized implementation of QPE can yield runtimes approaching those of state-of-the-art classical methods.
The presence of many opportunities for improving the quantum algorithm as well as tightening bounds on algorithm performance suggest however that quantum computers may ultimately realize a significant advantage for these instances.

\begin{figure}[ht]
    \centering
    \includegraphics[width=0.6\textwidth]{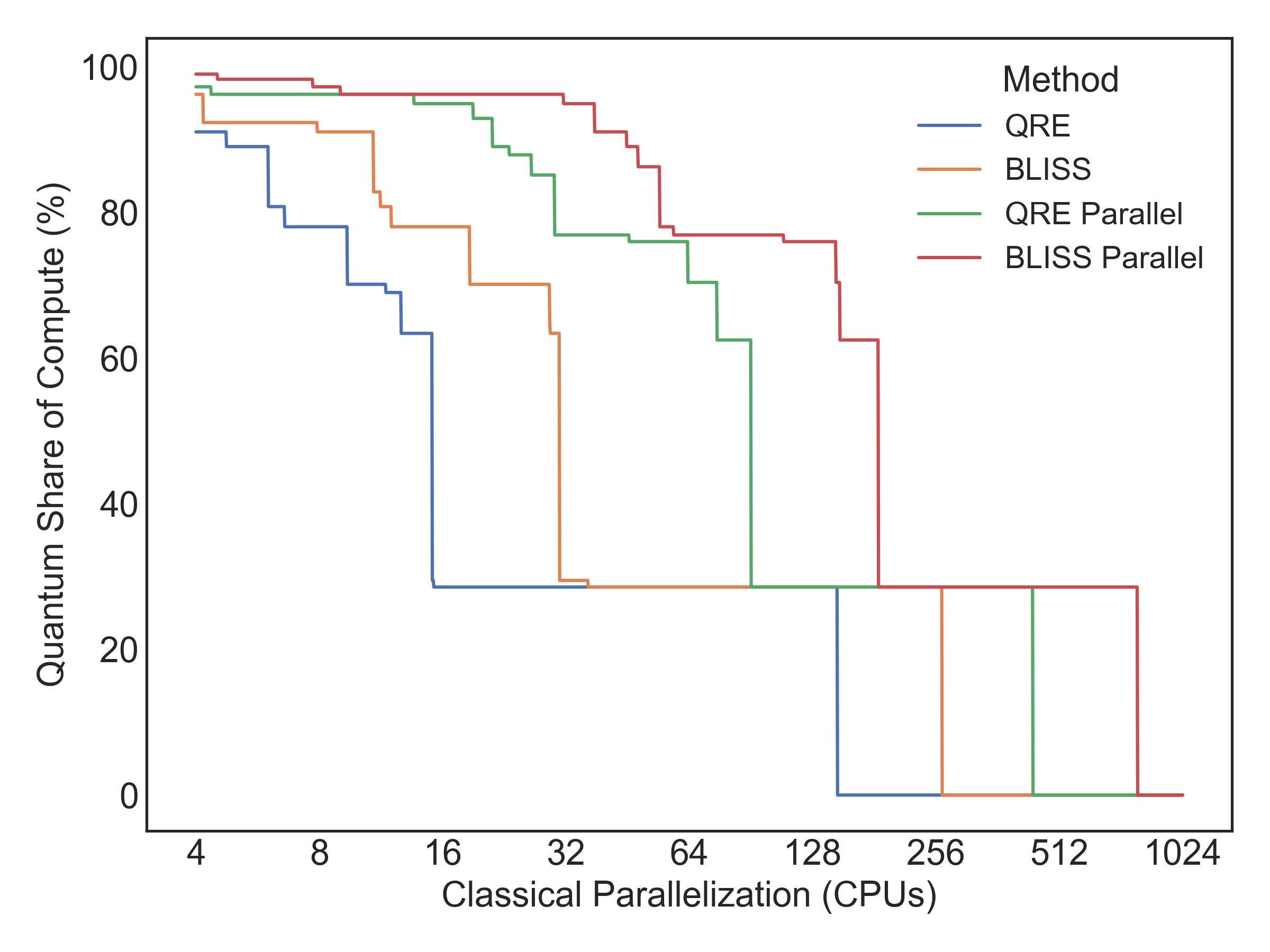}
    \caption{Quantum Share of Compute (\%) vs. Classical Parallelization (CPUs). This plot illustrates how increased classical parallelization can allow classical computers to outperform quantum methods even for large problems, highlighting the competitiveness of highly parallelized classical approaches.}
    \label{fig:quantum_share}
\end{figure}

Currently, classical methods can likely compete with our quantum resource estimates through high parallelization. This is demonstrated by the degree of parallelization needed to surpass quantum run times, as shown in Figure \ref{fig:quantum_share}. This figure illustrates the diminishing quantum share of compute as classical parallelization scales up. Although we assume perfect parallelization for the classical method, which may be unrealistic, \verb|Block2| has been shown to achieve results beyond what we explore here.

There are several reasons why the comparison between classical and quantum resources presented here is somewhat more pessimistic than that presented in related studies on molecular systems~\cite{goings2022,Lee2020,vonBurg2021}.
First, this work employs a stricter performance requirement (99\% probability of achieving an energy error less than 1.6~mHa when accounting for all error sources).
Prior studies permitted looser requirements, for example allowing a $\sim10\%$ failure rate per shot from hardware error, allowing a root-mean-square phase estimation error of 1~mHa, and neglecting costs associated with imperfect ground-state overlap
\cite{goings2022,Lee2020,vonBurg2021}.
(Note that raising the allowable failure rate from 1\%, but to 33\% reduces our estimates by an order of magnitude, as shown in Fig. \ref{fig:runtime-vs-failure-rate}.)
Second, some prior studies considered more performant quantum algorithms and more optimized algorithm parameters.
In particular, some prior studies considered tensor hypercontraction (THC) block encoding and also determined the number of bits of precision needed for encoding angles and coefficients empirically rather than using upper bounds~\cite{goings2022,Lee2020}.
Third, our analysis considers the possibility of parallelizing DMRG calculations across a greater number of cores (up to 512) than prior resource estimates (32)~\cite{goings2022}.

Integrating the algorithmic improvements such as those described above, as well as tightening bounds on algorithm performance, could allow for a significant reduction in the resource estimates presented here.
Potential algorithmic improvements include:
\begin{itemize}
    \item More efficient block encodings such as Tensor Hypercontraction (THC)~\cite{Lee2020} or Symmetry-Compressed Double Factorization (SCDF)~\cite{rocca2024reducing}.
    \item Initial states that yield higher overlaps, such as matrix product states~\cite {fomichev2023initial} or spin-coupled states~\cite{martidafcik2024spin}.
    \item First-quantized algorithms~\cite{su2021first}.
    \item More optimal choices for algorithm parameters.
\end{itemize}
Bounds used in the performance model that potentially may significantly overestimate errors include:
\begin{itemize}
    \item The Chebyshev bound on phase estimation error (Eq. \ref{eqn:prob-spectral-leakage-failure}).
    \item The approximate bound on the root-mean-square phase estimation error (Eq. \ref{eqn:spectral-leakage-variance})
    \item Bounds on the block encoding error (Eqs. \ref{eqn:coef-encoding-error} and \ref{eqn:angle-encoding-error})
    \item The bound on the error of the minimum sampled energy (Eq. \ref{eqn:total-qpe-failure-bound}).
\end{itemize}

We speculate that when such algorithm and performance model improvements are accounted for, quantum phase estimation may exhibit a significant advantage over classical methods.
For example, if each item listed above reduced the quantum runtime by a factor of two, then the overall runtime would be reduced by over two orders of magnitude.
Improvements to the classical methods suggested here are possible as well, although classical methods have already undergone extensive optimization historically.
Furthermore, the ability to directly benchmark classical algorithms obviates the need to resort to upper bounds on their performance.

In conclusion, continued investment in quantum algorithms is expected to enable quantum computing to surpass classical methods for specific complex problems.
We emphasize that because this study includes chemically diverse instances with varying charges, spins, and sizes, the results may apply not only to other problems in catalysis but a broad range of chemical systems of practical instance.
Our work illustrates the substantial potential of quantum computing in advancing catalytic science, provided that the technology continues to evolve to become more accessible, cost-effective, and capable of handling diverse chemical challenges.

\section{Acknowledgements}
This work was funded under the DARPA Quantum Benchmarking program. The authors wish to acknowledge the DARPA QB and Test \& Evaluation teams at NASA QuAIL and MIT-LL for helpful guidance and feedback. Special thanks to Michael Garrett and John Penuel at L3H Technologies for their guidance as Co-PIs of the DARPA Quantum Benchmarking program, which shaped the overall direction of this work. We also thank K. Morrell for valuable discussions and suggestions, R. Babbush for sharing insights regarding ground-state overlap and Hamiltonian factorizations, C. Gidney for elucidating nuances of modeling of magic state distillation, and D. Wellnitz for valuable discussions.
This research was partly enabled by Compute Ontario (computeontario.ca) and the Digital Research Alliance of Canada (alliancecan.ca) support. Part of the computations were performed on the Niagara supercomputer at the SciNet HPC Consortium. SciNet is funded by Innovation, Science, and Economic Development Canada, the Digital Research Alliance of Canada, the Ontario Research Fund: Research Excellence, and the University of Toronto. 
J.T.C. would like to thank A.M. Rey for her hospitality during his visit to JILA at CU Boulder. 

\appendix
\section{Quantum Resources Optimization: L1 Norm}
\label{app:lpbliss}
In this paper, we implement the qubitized Hamiltonian for quantum phase estimation using the linear combination of unitaries (LCU). The quantum resources required for LCU scale linearly (up to polylogarithmic terms) with the L1 norm $\lambda$ of the Hamiltonian \cite{lowHamSim2019,childsLCU2012,loaizaSimCost2023,loaiza2023}. The L1 norm is lower-bounded by the Hamiltonian spectral width $\Delta E$: $\lambda \geq \Delta E/2$ \cite{loaizaSimCost2023}. 
One significant factor contributing to the spectral width of a quantum-simulated electronic Hamiltonian is the increase in the accessible Fock space. That is, while the implemented Hamiltonian does not couple states with different particle numbers (i.e., different numbers of electrons), it still \textit{acts} on states of diverse particle number. 
These additional particle number subspaces increase the total spectral width of the Hamiltonian and thus  increase the lower bound on the L1 norm and the computational cost. 

The block-invariant symmetry shift method (BLISS)  \cite{loaiza2023} reduces the contribution to the spectral width of these extraneous particle number subspaces. BLISS adds operators $\hat K$ to the Hamiltonian $\hat H$  that modify the spectrum of the Hamiltonian, but
that do not impact the desired particle number subspace. That is, $\hat K\ket{\eta} = 0$, where $\ket{\eta}$ is any state within the desired particle number subspace. This results in a new Hamltonian $\hat H' = \hat H+ \hat K$ such that $\hat H'\ket{GS} = \hat H\ket{GS} =  E_{GS}\ket{GS}$, where $E_{GS}$ is the ground state energy and $\ket{GS}$ the ground state of the original Hamiltonian.
The $\hat K$ operators are  defined with multiple free parameters that are  varied to minimize the L1 norm of the target Hamiltonian. Overall, BLISS changes the Hamiltonian spectrum for states with the incorrect number of electrons, while preserving the Hamiltonian spectrum for states with the correct number of electrons. BLISS has been shown to reduce the L1 norm of several small molecules by a factor of about two \cite{loaiza2023} and has been made publicly available in the Julia package \href{https://github.com/iloaiza/QuantumMAMBO.jl}{\url{QuantumMAMBO}}.

We have developed a more efficient version, linear programming BLISS (LPBLISS), that replaces the non-linear gradient-descent-based optimization of BLISS with a linear programming approach that allows BLISS to be applied to larger systems. LPBLISS has been incorporated into a separate branch of QuantumMAMBO, found here \href{https://github.com/iloaiza/QuantumMAMBO.jl/tree/beta_merge}{\url{QuantumMAMBO:beta_merge}}.
Installation and running of LPBLISS has been designed to be straightforward and to take as input an FCIDUMP file describing a Hamiltonian. LPBLISS then outputs the modified Hamiltonian in another FCIDUMP file, making it ready for use in other parts of the quantum resource estimation pipeline.

We have results for the nitrogen fixation catalyst Hamiltonians discussed in Section \ref{sec:app_paper_prob_instances}, 
for both Pauli decomposition and double factorization (DF), in Table \ref{tab:lpbliss_results}. 
For the molecules shown, both LPBLISS and DF reduced the L1 norm with respect to the ``naive'' Pauli L1 norm with average factors of 1.4 and 2.8, respectively. If we assume that LPBLISS and DF work independently, we would expect an average improvement of 3.8. However, combining both LPBLISS and DF reduces the L1 norm by an average factor of 5.9, showing their synergistic improvement (see column $A/C$ of Table \ref{tab:lpbliss_results}). This synergistic improvement shows that LPBLISS is not only reducing the half spectral width ($\Delta E/2$) of the full Fock space (FS), but doing so in a way that allows DF to be even more effective than for the original Hamiltonian.

\begin{savenotes}
\begin{table}[!p]
    \setlength\tabcolsep{0pt}
    \centering
    \caption{
    LPBLISS results for catalyst Hamiltonians (see Section \ref{sec:app_paper_prob_instances}).
    The given L1 norms are for the Hamiltonian decomposed as a sum of Pauli products (Pauli) or decomposed via double factorization (DF).
    $N_o$ is the number of orbitals in the active space. 
    $\Delta E/2=(E_{max}-E_{min})/2$ is one half of the spectral width, calculated over the entire Fock space (FS HF) and the Electron Number Subspace
    (ENS HF)\cite{lpbliss_ens_note} using Hartree-Fock.
    $A/C$ and $B/C$ are the ratios of the labelled columns.} \label{tab:lpbliss_results}
    \begin{subtable}{\textwidth}
    \caption{LPBLISS results for Schrock catalyst (Section \ref{sec:schrock_mo_n2_instance})} \label{tab:lpbliss_mo_n2}
    \begin{tabular*}{\textwidth}{@{\extracolsep{\fill}}lcccccccccc}
        \toprule
        \multirow[10pt]{2}{*}{\shortstack{Molecule\\ID}}  & 
        \multirow{2}{*}{$N_o$} & \multicolumn{2}{c}{Original} &  
        \multicolumn{4}{c}{LPBLISS-treated} & 
        \multirow{2}{*}{\shortstack{A\\\textbf{---}\\C}}&
        \multirow{2}{*}{\shortstack{B\\\textbf{---}\\C}}&
        \multirow{2}{*}{\shortstack{LPBLISS \\Calculation \\Time (sec)}} \\ 
        \cmidrule(lr){3-4}\cmidrule(lr){5-8}
        &&
        \shortstack{(A)\\Pauli\\L1 Norm}&
        \shortstack{(B)\\DF L1\\Norm}&
        \shortstack{Pauli\\L1 Norm}&
        \shortstack{(C)\\DF L1\\Norm}&
        \shortstack{$\Delta E/2$,\\FS\\HF}&
        \shortstack{$\Delta E/2$,\\ENS\\HF}&&& \\
        \midrule 
        MoN$_2^-$ & 33 & 816.52 & 260.93 & 599.65 & 98.52 & 20.87 & 20.21 & 8.3 & 2.6 & 173.5\cite{lpbliss_calc_time_note} \\
        MoN$_2$ & 33 & 839.72 & 260.07 & 619.46 & 103.29 & 20.77 & 20.18 & 8.1 & 2.5 & 174.0\cite{lpbliss_calc_time_note} \\ 
        Fe(Cp)$_2^+$ & 46  & 1662.54 & 487.97 & 1284.03 & 250.38 & 26.04 & 25.75 & 6.6 & 2.0 & 1109.3\cite{lpbliss_calc_time_note} \\
        Fe(Cp)$_2$ & 46  & 1660.52 & 489.07 & 1275.49 & 247.99 & 26.05 & 26.03 & 6.7 & 2.0 & 1116.8\cite{lpbliss_calc_time_note} \\
        \bottomrule
    \end{tabular*}
    \end{subtable}

    \vspace*{2.5mm}

    \begin{subtable}{\textwidth}
    \caption{LPBLISS results for Bridged Dimolybdenum (Section \ref{sec:mo2_n2_instance})} \label{tab:lpbliss_mo2_n2}
    \begin{tabular*}{\textwidth}{@{\extracolsep{\fill}}lcccccccccc}
        \toprule
        \multirow[10pt]{2}{*}{\shortstack{Molecule\\ID}}  & 
        \multirow{2}{*}{$N_o$} & \multicolumn{2}{c}{Original} &  
        \multicolumn{4}{c}{LPBLISS-treated} & 
        \multirow{2}{*}{\shortstack{A\\\textbf{---}\\C}}&
        \multirow{2}{*}{\shortstack{B\\\textbf{---}\\C}}&
        \multirow{2}{*}{\shortstack{LPBLISS\\Calculation\\Time (sec)}} \\ 
        \cmidrule(lr){3-4}\cmidrule(lr){5-8}
        &&
        \shortstack{(A)\\Pauli\\L1 Norm}&
        \shortstack{(B)\\DF L1\\Norm}&
        \shortstack{Pauli\\L1 Norm}&
        \shortstack{(C)\\DF L1\\Norm}&
        \shortstack{$\Delta E/2$,\\FS\\HF}&
        \shortstack{$\Delta E/2$,\\ENS\\HF}&&& \\
        \midrule 
        1-Lut$_{Re}$ & 69  & 3166.70 & 901.92 & 2630.13 & 550.87 & 48.64 & 42.33 & 5.7 & 1.6 & 9350.1 \\ 
        1-Lut$_{TS}$ & 70  & 3526.25 & 903.83 & 2966.75 & 552.50 & 47.99 & 40.70 & 6.4 & 1.6 & 8889.2 \\ 
        II-Lut$_{Prod}$ & 70  & 3540.78 & 910.98 & 2978.27 & 547.76 & 48.18 & 40.35 & 6.5 & 1.7 & 3711.0 \\
        \bottomrule
    \end{tabular*}
    \end{subtable}

    \vspace*{2.5mm}

    \begin{subtable}{\textwidth}
    \caption{LPBLISS results for Mo Pincer reaction (Section \ref{sec:mo_n2_pincer_instance})}\label{tab:lpbliss_mo_n2_pincer}
    \begin{tabular*}{\textwidth}{@{\extracolsep{\fill}}lcccccccccc}
        \toprule
        \multirow[10pt]{2}{*}{\shortstack{Molecule\\ID}}  & 
        \multirow{2}{*}{$N_o$} & \multicolumn{2}{c}{Original} &  
        \multicolumn{4}{c}{LPBLISS-treated} & 
        \multirow{2}{*}{\shortstack{A\\\textbf{---}\\C}}&
        \multirow{2}{*}{\shortstack{B\\\textbf{---}\\C}}&
        \multirow{2}{*}{\shortstack{LPBLISS \\Calculation\\Time (sec)}}\\ 
        \cmidrule(lr){3-4}\cmidrule(lr){5-8}
        &&
        \shortstack{(A)\\Pauli\\L1 Norm}&
        \shortstack{(B)\\DF L1\\Norm}&
        \shortstack{Pauli\\L1 Norm}&
        \shortstack{(C)\\DF L1\\Norm}&
        \shortstack{$\Delta E/2$,\\FS\\HF}&
        \shortstack{$\Delta E/2$,\\ENS\\HF}&&& \\
        \midrule       
        \multicolumn{11}{c}{step (i) smaller active space} \\ \midrule
        RC & 32  & 513.13 & 318.05 & 291.95 & 117.75 & 61.20 & 58.16 & 4.4 & 2.7 & 142.2\cite{lpbliss_calc_time_note} \\ 
        TS$_{1/2}$ & 27  & 458.18 & 248.79 & 280.19 & 89.46 & 50.76 & 46.21 & 5.1 & 2.8 & 42.1\cite{lpbliss_calc_time_note} \\
        PC & 32  & 692.77 & 320.89 & 451.69 & 126.77 & 61.73 & 58.36 & 5.5 & 2.5 & 124.0\cite{lpbliss_calc_time_note} \\
        2 & 33  & 726.87 & 337.04 & 478.85 & 139.21 & 63.39 & 63.08 & 5.2 & 2.4 & 176.7\cite{lpbliss_calc_time_note} \\ 
        
        \midrule \multicolumn{11}{c}{step (i) larger active space} \\ \midrule
        RC & 51  & 1255.53 & 610.27 & 788.78 & 251.73 & 81.80 & 76.70 & 5.0 & 2.4 & 499.5 \\ 
        TS$_{1/2}$ & 51  & 1790.18 & 622.45 & 1267.93 & 278.54 & 77.73 & 77.62 & 6.4 & 2.2 & 535.3 \\
        PC & 51  & 1678.76 & 622.26 & 1168.17 & 271.84 & 79.28 & 77.70 & 6.2 & 2.3 & 496.1 \\
        2 & 52  & 1844.53 & 647.50 & 1292.61 & 292.16 & 81.06 & 79.12 & 6.3 & 2.2 & 503.9 \\ 
        
        \midrule \multicolumn{11}{c}{step (ii) smaller active space} \\ \midrule
        I & 56  & 2722.73 & 709.97 & 2276.01 & 410.07 & 101.53 & 93.00 & 6.6 & 1.7 & 2426.8 \\
        TS$_{\text{I/4a}}$ & 56  & 2119.43 & 664.56 & 1726.81 & 388.26 & 102.89 & 88.54 & 5.5 & 1.7 & 810.9 \\
        PC$^-$ & 55  & 1129.58 & 585.60 & 846.95 & 333.80 & 111.65 & 90.43 & 3.4 & 1.8 & 731.3 \\
        4a & 24  & 291.32 & 216.77 & 129.93 & 66.18 & 38.54 & 36.05 & 4.4 & 3.3 & 26.7\cite{lpbliss_calc_time_note} \\ 

        \midrule \multicolumn{11}{c}{step (ii) larger active space} \\ \midrule
        I & 75  & 4539.04 & 1119.83 & 3738.70 & 594.33 & 118.47 & 101.38 & 7.6 & 1.9 & 6078.4 \\ 
        TS$_{\text{I/4a}}$ & 73  & 3007.10 & 1030.02 & 2328.98 & 522.17 & 122.12 & 100.52 & 5.8 & 2.0 & 4993.2 \\
        PC$^-$ & 73  & 2079.47 & 954.69 & 1505.42 & 479.61 & 130.73 & 101.16 & 4.3 & 2.0 & 5343.4 \\ 
        4a & 43  & 873.40 & 453.05 & 500.16 & 189.48 & 68.59 & 67.56 & 4.6 & 2.4 & 776.9\cite{lpbliss_calc_time_note} \\ 
        \bottomrule
    \end{tabular*}
    \end{subtable}
\end{table}
\end{savenotes}

Also shown in Table \ref{tab:lpbliss_results} are the Hartree-Fock (HF) approximations to the half spectral width ($\Delta E/2$) for the full Fock space (FS) and the electron number subspace (ENS) of the specified number of electrons in the molecule. We can see that LPBLISS has been successful in reducing $\Delta E/2$ FS HF to within, on average, 9.4\% of the ENS bandwidth, with a maximum deviation of just 29.2\%. These numbers are in comparison to the average factor of 4 for the non-treated case (data not shown).

It appears that there is still some room for improvement from methods that do not modify the spectrum, as the DF L1 norm can be up to 11.5 times larger than its $\Delta E/2$ FS HF lower bound for the molecules shown (average factor of 4.9). On the other hand, as the full Fock space half spectral width is already close to that of the electron number subspace, any spectral-modifying methods would need to also modify the desired electron number subspace to obtain appreciable improvements on the half spectral bandwidth lower bound. 

Based on the ratio of the original DF L1 Norm to the LPBLISS-treated L1 Norm (column $B/C$ of Table \ref{tab:lpbliss_results}) and the linear dependence of the quantum runtime on the L1 norm \cite{lowHamSim2019,childsLCU2012,loaizaSimCost2023,loaiza2023}, we would expect an average improvement of 2.2 in the number of Toffoli gates required, assuming a prefactor of one.  As discussed in Sections \ref{sec:nitrogen_fixation_lre} and \ref{subsec:phys_qre}, we have performed physical and logical quantum resource estimates on the LPBLISS-treated Hamiltonians for the Schrock and Bridged Dimolybdenum catalysts; the results are in Tables \ref{tab:nitrogen_fixation_lre_lpbliss} and \ref{tab:nitrogen_fixation_pre_LPBLISS}. 

For the logical resource estimates of the Schrock catalyst, we see that the number of Toffoli gates dropped by a factor of about two. Unexpectedly, the number of logical qubits also dropped, by about 4\%. For the Bridged Dimolybdenum catalyst, both the number of Toffoli gates and the number of logical qubits dropped by a factor of about two.

For the physical resource estimates of the Schrock catalyst, we see that the runtime decreased by a factor of about two and the number of physical qubits decreased by about 6\%, with a mean spacetime volume reduction of 2.28. The space-time volume reductions are about the same as the reduction in the L1 norm induced by LPBLISS. For the Bridged Dimolybdenum catalyst, the reductions are more significant. The runtime again dropped by a factor of about 2, while the number of physical qubits dropped by a factor of about 20, leading to a mean space-time volume drop of about 37. This drop in space-time volume is about 23 times larger than the corresponding reduction in L1 norm. 

We thus see the level of improvement LPBLISS may have on quantum resources. However, given the large difference in improvement between these two systems and the small number of data points, it remains to be seen what  improvement (if any) could be expected from LPBLISS in general. The Pincer Hamiltonians are under investigation towards this end.
Detailed description of LPBLISS, analysis for more molecules and further analysis of the method will be performed in a separate manuscript \cite{2024lpbliss}.

\section{Algorithm Performance model}
\label{sec:algorithm-perf-model}

This section describes in detail the algorithm performance model and the methodology used for choosing algorithm parameters.
The probability $\delta$ that the lowest energy measured by QPE $E_\text{min}$ deviates from the true ground state energy of the active space Hamiltonian $E_0$ is bounded by
\begin{equation}
    \label{eqn:total-failure-bound}
    \delta \le 1 - \left(1 - \delta_\text{HW}\right)\left(1 - \delta_\text{GS}\right)\left(1 - \delta_\text{QPE}\right). 
\end{equation}
where $\delta_\text{HW}$ is the probability that a hardware error occurred in any of the $M$ shots; $\delta_\text{GS}$ is the probability that, in the absence of any hardware errors, none of the $M$ shots project into the ground state of the encoded Hamiltonian; and $\delta_\text{QPE}$ is the probability that the lowest energy measured by QPE $E_\text{min}$ deviates from the true ground state energy of the active space Hamiltonian $E_0$ by more than $\epsilon$, assuming that no shots experienced a hardware error and at least one shot projected into the ground state of the encoded Hamiltonian.
Note that the above expression is an inequality and not an equality because there is a possibility that the $E_\text{min}$ is within $\epsilon$ of $E_0$ if even one or more shots experiences a hardware failure or if none of the shots projects into the ground state of the encoded Hamiltonian.

The probability that one or more of the $M$ shots experienced a hardware failure is
\begin{equation}
    \label{eqn:hw-failure-bound}
    \delta_\text{HW} = 1 - \left(1 - p'_\text{HW}\right)^M
\end{equation}
where $p'_\text{HW}$ is the probability that an individual shot experiences a hardware failure.
Similarly, the probability that none of the $M$ shots projects onto the ground state of the encoded Hamiltonian is
\begin{equation}
    \label{eqn:projection-failure-bound}
    \delta_\text{GS} = \left(1-\left|\gamma\right|^2\right)^M
\end{equation}
where $\gamma = \braket{\psi}{\psi_0}$ is the overlap between the initial state $\ket{\psi}$ and the ground state of the encoded Hamiltonian $\ket{\psi_0}$.

Next, consider the probability $\delta_\text{QPE}$ that the lowest energy measured amongst the $M$ shots is not within $\epsilon$ of $E_0$ assuming that at least one shot projected into the ground state and that none of the shots experienced a hardware error.
Assuming that eigenstate projection error always yields states with the particle number of interest, this probability is bounded by the probability that all $M$ shots yield an energy within $\epsilon$ of the eigenvalue of the projected state.
This yields the expression
\begin{equation}
\label{eqn:total-qpe-failure-bound}
\delta_\text{QPE} \le 1 - \left( 1 -  p'_\text{QPE}\right)^M
\end{equation}
where $p'_\text{QPE}$ is the probability that an individual shot fails to yield an energy within $\epsilon$ of the eigenvalue of the state onto which it projected (which may or may not be the ground state).

There are two sources of algorithmic error which cause the measured energy to deviate from the eigenvalue of the projected state:
spectral leakage and block encoding error (i.e. the deviation between the eigenvalue of the projected eigenstate of the encoded Hamiltonian and the corresponding eigenvalue of the active space Hamiltonian).
Let $\sigma^2$ be the variance in the estimated energy arising from spectral leakage.
Then, from Chebyshev's inequality, the probability $\delta_\text{SL}$ that the spectral leakage error of a given shot exceeds some error threshold $\epsilon_\text{SL}$ is bounded by
\begin{equation}
\label{eqn:prob-spectral-leakage-failure}
\delta_\text{SL} \le \frac{\sigma^2}{\epsilon_\text{SL}^2}
\end{equation}
As discussed in Section IIIA of Ref.~\cite{Lee2020} and Section IIC of Ref.~\cite{babbush2018linear}, the variance of the spectral leakage within qubitized phase estimation using a half-sine window is bounded, under cetain approximations, by
\begin{equation}
\label{eqn:spectral-leakage-variance}
\sigma^2 \le \left(\frac{\pi \lambda}{2 \mathcal{I}}\right)^2.
\end{equation}
Combining Eqs. \ref{eqn:prob-spectral-leakage-failure}
 and \ref{eqn:spectral-leakage-variance} provides a bound on $\delta_\text{SL}$ in terms of $\mathcal{I}$:
\begin{equation}
\label{eqn:prob-spectral-leakage-failure-iterations} 
\delta_\text{SL} \le \left(\frac{\pi \lambda}{2 \mathcal{I} \epsilon_\text{SL}}\right)^2.
\end{equation}

The block encoding error is more challenging to rigorously bound.
In the case of the double factorized block encoding, the main sources of error arise from Hamiltonian coefficient encoding, rotation angle encoding, and Hamiltonian truncation~\cite{vonBurg2021}.
(In principle there also are errors arising from the synthesis of single-qubit rotations, but following the analysis of von Burg et al. we will neglect these~\cite{vonBurg2021}.)
Bounds on the spectral norm of the error in the encoded Hamiltonian due to the number of bits for encoding angles and coefficients can be used to choose a number of bits for encoding Hamiltonian coefficients and rotation angles~\cite{vonBurg2021,Lee2020}, and for the purposes of this analysis, these bounds are assumed to apply also to the error in the eigenvalues.
Strict bounds on truncation error are not available, and so prior studies have used the truncation error at the CCSD(T) level of theory as a proxy~\cite{Lee2020}.
This work assumes the absolute CCSD(T) truncation error to be an upper bound on the true absolute truncation error, i.e.
\begin{equation}
\epsilon_\text{trunc.} \le \left|E_\text{CCSD(T)}\left(t\right) - E_\text{CCSD(T)}\left(0\right)\right|
\end{equation}
where $E_\text{CCSD(T)}\left(t\right)$ is the CCSD(T) energy calculated for the Hamiltonian when truncated with threshold $t$ and $E_\text{CCSD(T)}\left(0\right)$ is the CCSD(T) energy calculated for the Hamiltonian without truncation.
This assumption is motivated by the fact that classical methods can be used to estimate the energy contribution from the truncated terms.
Therefore, when such corrections are used, the true truncation error likely can be reduced well below the CCSD(T) truncation error.

The absolute total shift in energy due to these block encoding errors $\epsilon_\text{BE}$ is bounded by the sum of the bounds on the individual contributions:
\begin{equation}
\label{eqn:block-encoding-error-bound}
\epsilon_\text{BE} \le \epsilon_\text{ang.} + \epsilon_\text{coef.} + \epsilon_\text{trunc.}
\end{equation}

Therefore the absolute total phase estimation error will be bounded by
\begin{equation}
  \label{eqn:qpe-error-bound}
  \epsilon_\text{QPE} \le \epsilon_\text{BE} + \epsilon_\text{SL}
\end{equation}
with a probability bounded by
\begin{equation}
    \label{eqn:qpe-failure-bound}
    \delta_\text{QPE} = \delta_\text{SL} \le \left(\frac{\pi \lambda}{2 \mathcal{I} \epsilon_\text{SL}}\right)^2.
\end{equation}

Eqs. \labelcref{eqn:total-failure-bound,eqn:hw-failure-bound,eqn:projection-failure-bound,eqn:total-qpe-failure-bound,eqn:coef-encoding-error,eqn:angle-encoding-error,eqn:block-encoding-error-bound,eqn:qpe-failure-bound,eqn:qpe-error-bound} along with the bounds on the spectral norm of the error in the encoded Hamiltonian~\cite{vonBurg2021,Lee2020} represent the algorithm performance model and provide a bound on the probability $p$ that the total error will exceed $\epsilon$ in terms of the algorithm parameters $m$, $M$, $\epsilon_\text{HW}$, $\aleph$, $\beth$, and $t$; the problem instance properties $N$, $\lambda$, and $\gamma$; and free parameters $\delta_\text{GS}$, $\delta_\text{QPE}$, $\delta_\text{HW}$, $\epsilon_\text{SL}$, $\epsilon_\text{ang.}$, $\epsilon_\text{coef.}$, and $\epsilon_\text{trunc.}$.

In principle, one could perform an optimization over these free parameters in order to minimize resource requirements.
However such an optimization is beyond the scope of this work, and instead the free parameters are chosen as follows.
First, let $\overline{p}_\text{QPE} = 0.8 p$ and $\overline{p}_\text{GS} = \overline{p}_\text{HW} = 0.1 \overline{p}$.
The number of shots $M$ is then chosen as 
\begin{equation}
\label{eq:num-shots}
M = \ceil*{\frac{\log \overline{p}_\text{GS}}{\log\left(1-\left|\gamma\right|^2\right)}},
\end{equation} which from Eq. \ref{eqn:projection-failure-bound} guarantees that
$\delta_\text{GS} \le \overline{p}_\text{GS}$.
The acceptable hardware failure rate per shot is chosen as
\begin{equation}
    \label{eqn:hw-failure-rate}
    \delta'_\text{HW} = \lceil 1 - \left(1 - \delta_\text{HW}\right)^{1/M}\rceil,
\end{equation} ensuring from Eq. \ref{eqn:hw-failure-bound} that $\delta_\text{HW} \le \overline{p}_\text{HW}$.

Next, let 
\begin{equation}
\overline{\epsilon}_\text{SL} = 0.8 \epsilon
\end{equation}
and
\begin{equation}
\overline{\epsilon}_\text{ang.} = \overline{\epsilon}_\text{coef.} = \overline{\epsilon}_\text{trunc.} = 0.066 \epsilon.
\end{equation}
The truncation threshold $t$ is chosen by scanning over a grid of candidate values and choosing the highest threshold for which
\begin{equation}
\left|E_\text{CCSD(T)}\left(t\right) - E_\text{CCSD(T)}\left(0\right)\right| \le \overline{\epsilon}_\text{trunc.},
\end{equation}
ensuring that $\epsilon_\text{trunc.} \le \overline{\epsilon}_\text{trunc.}$.
The bits of precision for coefficients and rotation angles are chosen as~\cite{vonBurg2021,Lee2020}
\begin{equation}
    \label{eqn:coef-encoding-error}
    \aleph = \left\lceil 2.5 + \log \left(\lambda / \overline{\epsilon}_\text{coef.}\right) \right\rceil,
\end{equation}
and
\begin{equation}
    \label{eqn:angle-encoding-error}
    \beth = \left\lceil 5.625 + \log \left( \lambda N / \overline{\epsilon}_\text{ang.} \right) \right\rceil,
\end{equation}
so that $\epsilon_\text{coef.} \le \overline\epsilon_\text{coef.}$,
assuming that the bound on the spectral norm of error in the encoded Hamiltonian applies to the error in encoded eigenvalues.

The number of phase estimation iterations is chosen as
\begin{equation}
\label{eq:qpe-iterations}
\mathcal{I} = \ceil*{\frac{\pi \lambda}{2 
\sqrt{1 - \left( 1 - \overline{p}_\text{QPE}\right)^{1/M}}
\overline{\epsilon}_\text{SL}}}
\end{equation}
ensuring from Eqs. \ref{eqn:total-qpe-failure-bound} and \ref{eqn:qpe-failure-bound} that $\delta_\text{QPE} \le \overline{p}_\text{QPE}$.

Altogether, the above parameter assignments ensure that \begin{equation}
\delta_\text{HW} + \delta_\text{QPE} + \delta_\text{GS} \le \overline{\delta}.
\end{equation}
Combining this with the union bound and Eq. \ref{eqn:total-failure-bound} shows that the probability of the energy error exceeding $\overline{\epsilon}$ is less than $\overline{\delta}$.

Note that $\overline{\delta}_\text{SL}$ has been chosen to be larger than $\overline{\delta}_\text{GS}$ and $\overline{\delta}_\text{HW}$ due to the higher costs associated with reducing the probability of spectral leakage failure compared to reducing the probability of other failure mechanisms.
Similarly, $\overline{\epsilon}_\text{SL}$ has been chosen to be larger than $\overline{\epsilon}_\text{ang.}$, $\overline{\epsilon}_\text{coef.}$, $\overline{\epsilon}_\text{trunc.}$.
\section{Logical Architecture Model}
\label{app:logical_architecture_model}
Here we describe the logical architecture modeling approach used to generate the physical resource estimates of Section \ref{subsec:phys_qre}.
The purpose of the logical architecture compilation is to 
translate an abstract logical circuit into a set of instructions that can be supported by components of a physical architecture, also known as a layout.
We will use the logical architecture model given by Gidney and Fowler~\cite{gidney2019flexible}. This model converts the abstract logical resource counts (Toffoli and T gate counts and logical qubit counts) into logical architecture compiled resource counts.

\begin{figure}[ht]
    \centering
    \includegraphics[width=0.5\linewidth]{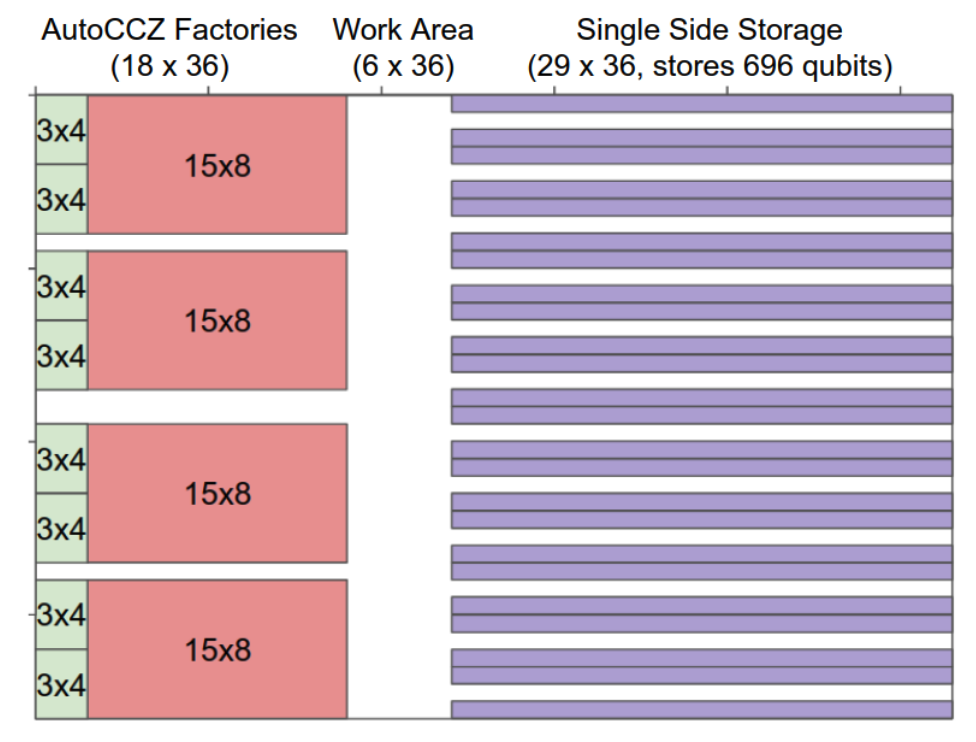}
    \caption{A diagram depicting the two-dimensional layout of the Fowler-Gidney architecture with figure borrowed from~\cite{Lee2020}. This shows the relative proportions of different usages of logical qubits in the architecture, tuned for a particular quantum chemistry application. Note that, instead of using the AutoCCZ factories, we have replaced these with Litinski factories~\cite{litinski2019magic} which are able to realize the lower magic state factory error rates required by the larger instances.}
    \label{fig:fowler-gidney-layout}
\end{figure}

Figure \ref{fig:fowler-gidney-layout} shows the layout of logical qubits used in this architecture as described in~\cite{Lee2020}.
The purple areas correspond to logical data qubits that store the wave function of the quantum computation. Four AutoCCZ factories, shown in red, run in parallel (with a green ``fixup'' region) and are used to simultaneously teleport Toffoli gates to the logical data qubits. The teleported Toffoli gates are ``routed'' through the white colored work area and along the ``hallways'' to reach the logical data qubits. For the large instances that we consider, the two-layer AutoCCZ factories do not achieve sufficiently low error rates. To address this, we replace the AutoCCZ factory models with Litinski factory models~\cite{litinski2019magic} that generate T gates and have comparable physical qubit footprints.
Accordingly, for each Toffoli gate in the original circuit, we account for four T gates~\cite{jones2013low}.

To model the performance of this logical architectures, several assumptions are made:
\begin{itemize}
    \item During the computation, the T gate production from the magic state factories is \emph{rate limiting}.
    \item During each round of magic state distillation, the work area qubits and data qubits either experience Clifford operations or must be ``kept alive'' through logical identity operations.
    \item The total logical space-time volume   of the quantum computation (i.e. the total number of elementary logical Clifford operations, including the identity operations) is the product of the number of logical qubits times the number of logical Clifford operations needed to implement all of the T gates for one factory (i.e. one quarter of that of the total number of T gates).
    \item The proportion of work area qubits (i.e. qubits used for routing) is 0.5 times the number of qubits used for storage. 
\end{itemize}
The number of physical qubits needed to support this logical architecture depends on the required logical error rate for each unit of space time volume.
We employ the logical error rate models that were used in~\cite{goings2022}, which take as input the code distance of the data qubits and the physical gate error rate, taken to be 0.001 for superconducting qubits.
From these architectural assumptions, the total runtime and physical qubit count are estimated as follows:
\begin{itemize}
    \item Physical qubit count: sum of the total number of physical qubits per factory times the number of factories plus the number of physical qubits for the work area plus the data qubits
    \item Runtime: product of the number of T gates gates and the time of each T gates gate divided by the number of parallel factories
\end{itemize}
To arrive at these numbers, we need to know the number of physical qubits per logical qubit for the factories and for the work area and data qubits. 
The number of physical qubits needed for each logical qubit is computed as follows: we loop over factories and over code distances for the non-factory qubits and find the minimal ``volume'' quantum computation such that the total failure probability is below $p_\text{QPE}$ (see Appendix~\ref{sec:algorithm-perf-model}).
This ``volume'' is computed as the product of the number of physical qubits and the total runtime.

\printbibliography

\end{document}